\documentclass[usenatbib]{mn2e}
\usepackage[toc,page]{appendix}
\usepackage{amsmath}
\usepackage[dvips]{graphicx}
\usepackage{epsfig}
\usepackage{lscape}
\usepackage{natbib}
\usepackage{amssymb}
\usepackage{amsmath}

\newcommand{\kms}{{\rm \,km\,s^{-1}}}
\newcommand{\Mpc}{\,{\rm Mpc}}
\newcommand{\kpc}{\,{\rm kpc}}

\newcommand{\Rcusp}{R_{\rm cusp}}
\newcommand{\Rfold}{R_{\rm fold}}

\newcommand{\Rein}{\theta_{\rm Ein}}

\title[Can CDM substructures fully explain radio flux-ratio anomalies]
      {How well can cold-dark-matter substructures account for the
        observed radio flux-ratio anomalies?}

\author[Xu et al.] {Dandan Xu$^{1}$\thanks{E-mail:
    Dandan.Xu@h-its.org}, Dominique Sluse$^{2,~3}$, Liang
  Gao$^{4,~5,~6}$, Jie Wang$^{4,~6}$, Carlos Frenk$^{5}$, \and Shude
  Mao$^{4,~6,~7,~8}$, Peter Schneider$^{3}$, Volker Springel$^{1,~9}$
  \\$^{1}$Heidelberg Institute for Theoretical Studies,
  Schloss-Wolfsbrunnenweg 35, 69118 Heidelberg, Germany
  \\$^{2}$Institut d'Astrophysique et de G\'eophysique, Universit\'e
  de Li\`ege, All\'ee du 6 Ao\^ut 17, B5c, 4000 Li\`ege, Belgium
  \\$^{3}$Argelander-Institut f$\ddot{u}$r Astronomie,
  Universit$\ddot{a}$t Bonn, Auf dem H$\ddot{u}$gel 71, 53121 Bonn,
  Germany \\$^{4}$National Astronomical Observatories, Chinese Academy
  of Sciences, 100012 Beijing, China \\$^{5}$Institute for
  Computational Cosmology, Dept. of Physics, University of Durham,
  South Road, Durham DH1 3LE, United Kingdom \\$^{6}$Key Laboratory
  for Computational Astrophysics, National Astronomical Observatories,
  Chinese Academy of Sciences, 100012 Beijing, China \\$^{7}$Jodrell
  Bank Centre for Astrophysics, the University of Manchester, Alan
  Turing Building, Manchester M13 9PL, United Kingdom \\$^{8}$Physics
  department and Tsinghua Center for Astrophysics, Tsinghua
  University, 100084, Beijing, China \\$^{9}$Zentrum f$\ddot{u}$r
  Astronomie der Universit$\ddot{a}$t Heidelberg, ARI,
  M$\ddot{o}$nchhofstr. 12-14, 69120 Heidelberg, Germany }

\date{Accepted ...... Received ...... ; in original form......   }

\begin{document}
\pagerange{\pageref{firstpage}--\pageref{lastpage}} \pubyear{2014}
\maketitle
\label{firstpage}
\begin{abstract}
Discrepancies between the observed and model-predicted radio flux
ratios are seen in a number of quadruply-lensed quasars. The most
favoured interpretation of these anomalies is that CDM substructures
present in lensing galaxies perturb the lens potentials and alter
image magnifications and thus flux ratios. So far no consensus has
emerged regarding whether or not the predicted CDM substructure
abundance fully accounts for the lensing flux anomaly
observations. Accurate modelling relies on a realistic lens sample in
terms of both the lens environment and internal structures and
substructures. In this paper we construct samples of generalised and
specific lens potentials, to which we add (rescaled) subhalo
populations from the galaxy-scale Aquarius and the cluster-scale
Phoenix simulation suites. We further investigate the lensing effects
from subhalos of masses several orders of magnitude below the
simulation resolution limit. The resulting flux ratio distributions
are compared to the currently best available sample of radio lenses.
%The flux ratios for systems with image triplets/pairs that are located
%further away from the critical curves (and thus have larger
%separations) are less susceptible to density fluctuations.  
The observed anomalies in B0128+437, B0712+472 and B1555+375 are more
likely to be caused by propagation effects or oversimplified/improper
lens modelling, signs of which are already seen in the data. Among the
quadruple systems that have closely located image triplets/pairs, the
anomalous flux ratios of MG0414+0534 can be reproduced by adding CDM
subhalos to its macroscopic lens potential, with a probability of
$5\%-20\%$; for B0712+472, B1422+231, B1555+375 and B2045+265, these
probabilities are only of a few percent. We hence find that CDM
substructures are unlikely to be the whole reason for radio flux
anomalies. We discuss other possible effects that might also be at
work.
%Again simplified lens modelling could have affected the flux ratio
%predictions; caveats can be traced in the lens modelling for B0712+472
%and B1555+375 (add B1422). Furthermore, baryonic substructures may
%also play an important role in causing density perturbation and thus
%help to bridge the gap between observations and predictions.

\end{abstract}


\section{Introduction}
Understanding the radio flux ratios of multiply-imaged quasars has
been a long-standing problem. In these systems, standard parametric
models of the lens mass distribution (e.g., a singular isothermal
ellipsoid plus external shear, hereafter ``SIE+$\gamma$'') can fit the
image positions well, but not their flux ratios. This is known as the
``anomalous flux ratio'' problem (\citealt{Kochanek91}).

A number of solutions have been proposed.  For example, some of the
flux-ratio anomalies could be accommodated by adding higher order
multipoles to the ellipsoidal potential of the lensing galaxy. However,
the required amplitudes are deemed to be unreasonably larger than
typically observed in galaxies and halo models
(\citealt{EW2003,KD2004,CongdonKeeton2005,Yoo2006}).

Propagation effects in the interstellar medium, such as galactic
scintillation and scatter broadening, could also cause anomalous flux
ratios. If so, one would expect a strong wavelength dependence of the
anomalies measured at radio wavelengths, which was not seen
(\citealt{Koopmans2003JVASCLASS,KD2004}). Moreover, neither of the two
solutions proposed above could explain the observed parity dependence
of the flux anomalies (e.g., Metcalf \& Madau 2001;
\citealt{SW2002apj,Keeton2003SaddleImages,KD2004}).

Currently the most favoured explanation of the radio flux-ratio
anomalies invokes the perturbation effects from small-scale structures
hosted by lensing galaxies. In the cold dark matter (CDM) model of
structure formation a large population of dark matter subhalos is
predicted to survive inside larger ``host'' halos. In galaxies like
the Milky Way, their number vastly exceeds the number of observed
satellites (about two dozen have been discovered in the Milky Way to
date). On the one hand, part of this discrepancy can be readily
understood on the basis of standard ideas, e.g., photo reionization
and stellar feedback on galaxy formation
\citep{Bullock2000ReionizationFixSatelliteCount,
  Benson2002PhotoionizationFixSatelliteCount,
  Bovill2009MissingSatellites,Cooper2010GalaxyCDM,
  Font2011GalaxySatelliteCDM, Guo2011Dwarf2cDgalaxyCDM}.
%In particular, the model of
%\citet{Benson2002PhotoionizationFixSatelliteCount} predicted a
%population of ultrafaint satellites, which was subsequently discovered
%in the Sloan Digital Sky Survey (SDSS)
The gap was also narrowed by the discovery of a population of
ultrafaint satellites from the Sloan Digital Sky Survey (SDSS)
\citep{Tollerud2008,Koposov2008MWSatelliteLF,
  Koposov2009MWSatellites}. Despite this, several controversies still
exist on small scales regarding the abundance (``missing satellite''
problem, e.g.,
\citealt{Klypin1999apj,Moore1999apj,Kravtsov2004MissingSatellites})
and the density profiles (``the cusp/core'' problem, see
\citealt{Ludlow2013pofile}) of these dark matter subhalos that are
predicted to exist but somehow failed to make galaxies.

If they do exist as CDM predicted, they could then be probed through
their gravitational lensing effects. Earlier studies from e.g.,
\citet{MS1998mn}, \citet{MM2001} and \citet{MZ2002}, proposed that
substructures (on scales much smaller than image separations of
$1\arcsec$ for typical lens and source redshifts) could explain the
radio flux-ratio anomalies in quadruply-lensed quasar images. Later
studies showed that the presence of substructures in lensing galaxies
can also explain the observed tendency for the brightness of the
saddle image to be suppressed (\citealt{SW2002apj,
  Keeton2003SaddleImages, KD2004}). The perturbations by subhalos have
therefore emerged as one of the most convincing explanations for the
radio flux-ratio anomalies. If true, such an explanation could have
important implications for cosmology since it provides a direct and
crucial test of the CDM model.

To date, there are about a dozen studies that use $N$-body simulations
to test if the predicted CDM substructures have the right amount to
explain the observed frequency of anomalous lenses in currently
available samples. However, no consensus has emerged. While some of
the studies (e.g., \citealt{DK2002, BS2004aa, DoblerKeeton2005,
  Metcalf2010FluxAnomaly}) suggest consistency between the CDM model
and observations, others (e.g., \citealt{MaoJing04apj, AB06mn,
  Maccio2006b, Maccio2006, Chen2011CuspViolation}) including those by
us (\citealt{Dandan09AquI,Dandan2010AqII}) find that subhalos from CDM
simulations are actually {\it insufficient} to explain the observed
radio flux-anomaly frequency.

To tackle this problem from the numerical simulation point of view,
one needs to model a realistic sample of the lens population, from
their larger-scale environment to their internal structures and
substructures. Any numerical experiment in this regard is facing
several major issues that directly affect the accuracy of the
study. For example, as shown by \citet{KGP2003apj}, flux ratios are
quite sensitive to the ellipticity of the main
lens. \citet{Metcalf2010FluxAnomaly} also pointed out that one of the
reasons that our previous studies
(\citealt{Dandan09AquI,Dandan2010AqII}) did not reproduce enough
perturbations to match observations could be due to our adoption of a
restricted ellipticity instead of the full range of ellipticities in
the main lens models.

Second, the lack of a proper subhalo population may have distorted our
previous conclusion. Previously we exclusively used the Milky
Way-sized halos from the Aquarius project
(\citealt{volker08Aq}). However, massive elliptical galaxies, which
comprise 80\%-90\% of observed lenses \citep{KKF1998,KFI2000,RKF2003}
are more likely to occur in group-sized halos which are generally ten
times more massive. Since the subhalo abundance increases rapidly with
increasing host halo mass \citep[e.g.][]{DeLucia2004CDMSub, Gao2004b,
  Zentner2005SubhaloPopulation, Wang2012MassiveHostMoreSub}, the
adoption of subhalo populations hosted by Milky Way-sized halos could
underestimate the probability of flux-ratio anomalies.

Third, at present, even the best cosmological $N$-body simulations
only resolve subhalos down to $10^{6\sim7}h^{-1}M_{\odot}$ so the
lensing effects from subhalos of masses beyond such resolution limit
cannot be readily studied using these $N$-body simulations. In the
cold dark matter cosmogony, low mass subhalos are much more abundant
than their higher mass counterparts. Should we expect more
perturbation effects from the low mass subhalos? Or what could be the
observational signatures of these substructures predicted to exist at
the lower levels of the hierarchy of cosmic structures?  Specifically,
down to which mass levels would the dark matter subhalos still be able
to affect the image brightness and flux ratios at radio wavelengths?

Last but not least, the cosmological simulations that have been used
in these studies contain only dark matter but no baryons, the
inclusion of which might change the subhalo survival rate as well as
their density profiles/concentration, that in turn might lead to a
different conclusion.

In this paper we accommodate the first three issues above and find
that for systems with image triplets/pairs of larger separation, whose
flux ratios are less susceptible to density fluctuations, their
observed anomalies are more likely to be caused by propagation effects
or simplified lens modelling; for systems with closely located
triplets/pairs, CDM substructures alone can only account for the
observed flux ratios with percent-level probabilities; therefore they
may not be the entire reason. We point out that other possible
sources, e.g., inadequate lens modelling again, as well as baryonic
substructures may also be at work. To this end, high resolution
hydrodynamic simulations are in great need to help us identifying
other possible culprits for the radio flux-ratio anomalies.

This paper is organized as follows: in the first part, we show that
using generalised lens models and simulated subhalo populations in
group-sized halos will indeed increase the flux anomaly
frequency. Specifically, in Sect. 2 we review the generic relations in
cusp (Sect. 2.1) and fold (Sect. 2.2) lenses, and present our
observational sample of eight systems, all of which have radio
measurements for both cusp and fold relations (Sect. 2.3). In Sect. 3,
we present the method to model massive elliptical lenses and their
subhalo populations. For the former (in Sect. 3.1), we use a technique
similar to that of Keeton et al. (2003). For the latter (in
Sect. 3.2), we rescale the subhalo populations from two sets of
high-resolution cosmological CDM simulations - the Aquarius
\citep{volker08Aq} and Phoenix \citep{Gao12Phoenix}, and add them to
the smooth lens potentials. 
%The resulting flux-ratio probability
%ldistributions are presented in Sect. 3.4.

In the second part of this paper, i.e., in Sect. 4, we focus on
individual observed systems, taking the best-fitting macroscopic lens
models and populate not only the rescaled Aquarius and Phoenix subhalo
populations above $10^7h^{-1}M_{\odot}$ (in Sect. 4.2) but also a
low-mass subhalo population down to masses two orders of magnitudes
below (in Sect. 4.3 and 4.4). The observational signatures of very
low mass subhalos and the dependence on source sizes are also studied
and results are presented in Sect. 4.5. The probabilities to reproduce
the observed flux ratios are given in Sect. 4.6. Finally a discussion
and our final conclusions are given in Sect. 5.

The cosmology we adopt here is the same as that for both sets of
simulations that we use in this work, with a matter density
$\Omega_{\rm m}$ = 0.25, cosmological constant $\Omega_{\Lambda}$ =
0.75, Hubble constant $h=H_0/(100\kms\Mpc^{-1})=0.73$ and linear
fluctuation amplitude $\sigma_8=0.9$. These values are consistent with
cosmological constraints from the WMAP 1- and 5-year data analyses
(\citealt{WMAP-1, WMAP-5}), but differ from the Planck 2013 results
(\citealt{Planck2013}), where $h=0.67$ and $\sigma_8=0.83$. We do not
expect these differences in cosmological parameters to have
significant consequences for our conclusions.

\section{Generic relations in cusp lenses and fold lenses}

\begin{figure}
\centering
\includegraphics[width=7cm]{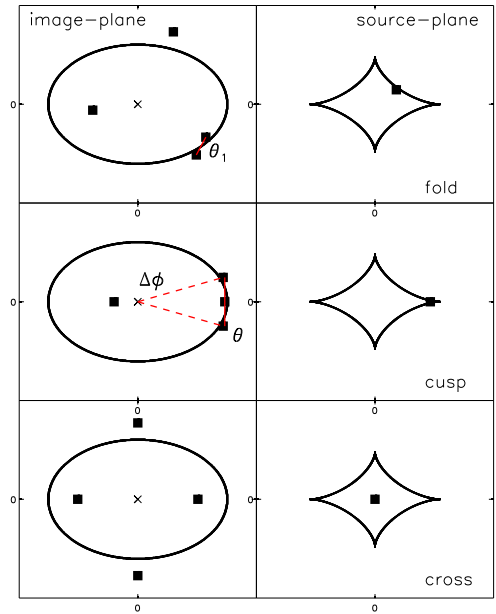}
\caption{Three basic image configurations: fold (top), cusp
(middle), and cross (bottom), with respect to the tangential
critical curves in the image plane (on the left), and corresponding
source positions with respect to the central caustics in the source
plane (on the right). The image separation $\theta_1$ of a close
pair is labelled for the fold configuration; image opening angle
$\Delta\phi$ and separation $\theta$ of a close triplet are labelled
for the cusp configuration.} \label{fig:CuspFoldCrossConfiguration}
\end{figure}

There are three generic configurations of four-image lenses (see
Fig. \ref{fig:CuspFoldCrossConfiguration}): (1) a source located near
a cusp of the tangential caustic will produce a ``cusp''
configuration, where three images form close to each other around
the critical curve on one side of the lens; (2) a source located
near the caustic and between two adjacent cusps will produce a
``fold'' configuration, where a pair of images form close to each
other near the critical curve; (3) a source located far away from
the caustic, i.e., in the central region of the caustic, will
produce a ``cross'' configuration, where all four images form far
away from each other and away from the critical curve. Close triple
images in cusp lenses and close pair images in fold lenses are the
brightest images among the four, as they form close to the
(tangential) critical curve.

There are some universal magnification relations for the triple and
pair images in cusp and fold systems in smooth lens
potentials. Without detailed lens modelling for individual systems,
these generic relations assist one in identifying small-scale
perturbations, which cause violations of these generic magnification
relations.

\subsection{The cusp relation}

In any smooth lens potential that produces multiple images (of a
single source) of a cusp configuration, a specific magnification ratio
(i.e., also flux ratio) of the image triplet will approach zero
asymptotically, as the source approaches a cusp of the tangential
caustic. This is known as the ``cusp relation'' (\citealt{BN1986apj,
  Mao92, SW1992aa, Zakharov1995AA, KGP2003apj}), mathematically defined
as:
\begin{equation}
  \Rcusp \equiv \frac{\mu_A + \mu_B +
    \mu_C}{|\mu_A|+|\mu_B|+|\mu_C|} \rightarrow 0 ~~~
  (\Delta\beta \rightarrow 0),
\label{eq:Rcusp}
\end{equation}
where $\Delta\beta$ is the offset between the source and the nearest
cusp of the caustic, $\mu_{A, B, C}$ denote the triplet's
magnifications, whose signs indicate image parities.

Because $\Delta\beta$ cannot be directly measured, we therefore
follow the practice of Keeton et al. (2003), using $\Delta\phi$ and
$\theta/\Rein$ to quantify a cusp image configuration. As labelled in
Fig. \ref{fig:CuspFoldCrossConfiguration}, $\Delta\phi$ is defined as
the angle between the outer two images of a triplet, measured from
the position of the lens centre; $\theta/\Rein$ is the maximum image
separation among the triplet, normalized by the Einstein radius
$\Rein$. In general, when the source moves towards the nearest cusp,
both $\Delta\phi$ and $\theta/\Rein$ will decrease to zero.

In particular small-scale structures, either within the lens or
projected by chance along the line of sight, will perturb the lens
potential and alter fluxes of one or more images. In this case,
$\Rcusp$ will become unexpectedly large. The cusp relation, i.e.,
$\Rcusp \rightarrow 0$ when $\Delta\beta\rightarrow 0$, will then be
violated.

\subsection{The fold relation}

For an image pair in a fold configuration produced by any smooth lens
potential, there is also a generic magnification relation, namely the
``fold relation'' (\citealt{BN1986apj, SW1992aa, SchneiderBook1992,
  PettersBook2001}). In this paper, we take the form as in Keeton et
al. (2005):
\begin{equation}
  \Rfold \equiv \frac{\mu_{\rm min}+\mu_{\rm sad}}{|\mu_{\rm min}|+|\mu_{\rm sad}|}
  \rightarrow 0 ~~~ (\Delta\beta \rightarrow 0),
\label{eq:Rfold}
\end{equation}
where $\Delta\beta$ is the offset of the source from the fold
caustic, $\mu_{\rm min, sad}$ denote magnifications of the minimum
($\mu>0$) and saddle ($\mu<0$) images. To quantify a fold image
configuration, similar to the practice of Keeton et al. (2005), we
use $\theta_1/\Rein$ to indicate how close the pair of images are.
As labelled in Fig. \ref{fig:CuspFoldCrossConfiguration},
$\theta_1/\Rein$ is defined as the separation, in unit of the
Einstein radius $\Rein$, between the saddle image and the nearest
minimum image.

Once again, when small-scale structures are present, $\Rfold$ will
also become unexpectedly large; the fold relation, i.e., $\Rfold
\rightarrow 0$ when $\Delta\beta\rightarrow 0$, will then be violated.
In principle one can study the perturbing small-scale structures by
investigating the violations of the cusp and fold relations in extreme
systems where $\Delta\beta\sim0$. However, the detection of such
systems is rare. For observed lenses, $\Delta\beta\neq 0$; and the
exact values of $\Rcusp$ and $\Rfold$ depend on $\Delta\beta$, as well
as the lens potentials. Without detailed lens modelling, one can
identify cases of violations as outliers of some general distributions
of $\Rcusp$ and $\Rfold$ for smooth lenses. A series of comprehensive
and detailed studies on this topic have been carried out by e.g.,
Keeton et al. (2003, 2005), whose methods are largely followed in this
work (Sect. 3.1).

\subsection{A sample of cusp and fold lenses}

%There are more than twenty lenses with $\Rcusp$ and $\Rfold$ measured
%at optical, near-infrared (NIR) and radio wavelengths (CfA-Arizona
%Space Telescope Lens Survey,
%http://cfa-www.harvard.edu/castles). Nearly half of them show
%anomalous flux ratios in the sense that the measured $\Rcusp$ and
%$\Rfold$ cannot be reproduced by the smooth lens models that best fit
%their image positions (Keeton et al. 2003, 2005).

In order to quantify how well the CDM substructures can account for
the flux-anomaly observations, we take all the quadruple systems with
$\Rcusp$ and $\Rfold$ measured at radio wavelengths, as the fluxes
measured at optical and NIR (\citealt{SluseMIR2013}) wavelengths can
be significantly affected by stellar microlensing and dust extinction.
This forms a sample of a total of eight lenses; three ``cusp'', and
five ``fold'' lenses. The radio flux ratios of several other systems
can also be found in literature but are excluded from this work: three
systems have atypical nature of the lensing galaxy, i.e., the Einstein
cross Q2237+0305 (\citealt{Falco1996Q2237}) which is lensed by the
bulge of a spiral galaxy, the large separation system J1004+4112
(\citealt{Jackson2011J1004}) which is lensed by a galaxy cluster, and
B1359+154 (\citealt{Rusin2001B1359}) which shows six images and is
lensed by a group of galaxies; we also excluded MG2016+112
(\citealt{Garrett1994MG2016}) which is only triply imaged at radio
wavelengths.

To quantify the image geometry we take the basic image configuration
measurements, namely, $\Delta\phi$, $\theta/\Rein$ and
$\theta_1/\Rein$, as well as the measured and model-predicted flux
ratios of $\Rcusp$ (for the closest triple images) and $\Rfold$ (for
the closest saddle-minimum image pairs), as listed in Table
\ref{tab:ObsSample-Cusp}. It can be seen that discrepancies at
different levels exist between the measured flux ratios and model
predictions. Below we give a brief description of each individual
system in our lens sample.

\subsubsection{B0128+437}
This is a fold system. The observed flux ratios are likely affected by
complex systematic errors, as suggested by radio-frequency dependent
flux ratios and by VLBI imaging. The VLBI data show that the source is
composed of three aligned components, one being tentatively associated
with a flat spectrum core and the other two with steep spectrum
components of the jet. The lensed image $B$ only barely shows the
``triple'' structures, which are visible in images $A$, $C$ and
$D$. Hence it is likely that image $B$ is affected by scatter
broadening (Biggs et al. 2004). On the other hand, lens modelling using
the VLBI data suggests astrometric perturbations of image positions by
substructures (Biggs et al. 2004).

\subsubsection{MG0414+0534}
This is a fold system with a pair of images very close to the critical
curve (image magnifications $|\mu|>15$, see the lens modelling in
Sect. 4). The low-resolution radio observations of Lawrence et
al. (1995) lead to roughly the same $\Rcusp$ at multiple epochs and at
different frequencies with the VLA, suggesting that the time delay
between the lensed images is not a concern. However, a lower value of
$\Rcusp$ was obtained from higher-resolution VLBI observations of Ros
et al. (2000), which resolved the core+jet components of the
source. The ratios for the core images also agree well with the one
measured in MIR (\citealt{Minezaki2009MG0414}). The $\Rfold$ values
from VLA, VLBI, MIR and extinction-corrected optical data all agree
with each other within the measurement uncertainties. We use the VLBI
results of both $\Rcusp$ and $\Rfold$ (for the core images) in our
analysis.

\subsubsection{B0712+472}
This is a cusp/fold system with a close image configuration of
$\Delta\phi=76.9^{\circ}$. We use VLA flux ratios obtained at 5GHz by
Koopmans et al. (2003). Those ratios were observed to be stable over
41 epochs of monitoring spanning 8.5 months, and are compatible with
VLBI 5Ghz measurements (Jackson et al. 2000). The flux ratios deviate
significantly from the optical/NIR flux ratios, which are affected by
differential extinction and microlensing (Jackson et al. 1998, 2000).

\subsubsection{B1422+231}
This is a classical cusp lens with $\Delta\phi=77^{\circ}$. The flux
ratios taken here were measured at different radio frequencies, at
different epochs and with different spatial resolutions (with VLA and
VLBA), which all agree with each other, as well as with mid-infrared
(MIR) data (\citealt{Patnaik1992B1422, Patnaik1999B1422,
  Koopmans2003JVASCLASS, Chiba2005}). Recently, with the aid of the
adaptive optics integral field spectrograph on the Keck I Telescope,
\citet{Nierenberg2014B1422} derived the narrow-line flux ratios, which
are also consistent with those measured in the radio.

\subsubsection{B1555+375}
This is a fold system, with a pair of images predicted to be very
close to the critical curve (image magnifications $|\mu|>50$). The
radio fluxes were obtained at 5GHz with the VLA and averaged over 41
epochs over 8.5 months (Koopmans et al. 2003). The HST images of this
system also suggest that it is a very flattened lens.

\subsubsection{B1608+656} 
This is a two-lens system, and has a fold image configuration with a
relatively large opening angle. Lens models suggest that the image
magnifications are small ($|\mu|<5$). Many VLA data are available
(including monitoring data) for this system and show consistently
$\Rfold \sim 0.32$. The radio measurements of $\Rcusp$ and $\Rfold$
are larger than observed in the optical and NIR, where the source
appears to be extended and significantly affected by differential
extinction (\citealt{Surpi2003B1608}).

\subsubsection{B1933+503} 
This is also a fold system and lens models suggest that the image
magnifications are small ($|\mu|<5$). The VLBI images presented in
\citet{Suyu2012B1933} reveal that the cores in images 1 and 4 show two
peaks but not for image 3. This suggests that scatter broadening may
modify the radio flux ratios. The $\Rcusp$ and $\Rfold$ obtained from
this high resolution images also agree with lower resolution VLA and
MERLIN data (Sykes et al. 1998), which are used here.

\subsubsection{B2045+265}
This is a very extreme cusp lens with $\Delta\phi=34.9^{\circ}$. All
three images are located (symmetrically) close to the critical curve
with image magnifications $|\mu|>50$. The radio flux ratios are very
robust at different spatial resolution (VLA, VLBA) over different
periods of time, and consistent with the H-K wavelengths (Fassnacht et
al. 1999; McKean et al. 2007). Koopmans et al. (2003) identified
significant intrinsic variability at radio wavelengths, but the
amplitude of this effect is apparently small, at least on a time scale
of months. The VLBA data reveals a core+jet emission for image $A$,
but not for the saddle point image $B$, which should be brighter than
$A$ according to the models.
% This indicates a possibility for the presence of a substructure
% around image $B$ that demagnifies both the compact core and jet
% emissions.

\begin{table*}
\centering \caption{Observed lenses with measurements of $\Rcusp$
and $\Rfold$ for the close triple images:}
\label{tab:ObsSample-Cusp}
\begin{minipage} {\textwidth}
\begin{tabular}[b]{l|c|c|c|c|c|c|c}\hline\hline
  ~~Lens & ~~Type~~& ~~$\Delta\phi$(${^\circ}$)~~ & ~~$\theta/\Rein$~~ &
  ~~$\Rcusp$~~& ~~$\theta_1/\Rein$~~ & ~~$\Rfold$~~
  & ~References~ \\\hline
B0128+437$^\dagger$     & fold  & 123.3 & 1.511  & $-$0.043$\pm$0.020 ($-$0.090) & 0.584 & 0.263$\pm$0.014 (0.161) & 1, 2 \\
MG0414+0534   & fold  & 101.5 & 1.841  & 0.213$\pm$0.049 (0.118)  & 0.388 & 0.087$\pm$0.065 ($-$0.029) & 3, 4, 5, 6 \\
B0712+472 & cusp & 76.9  & 1.503  & 0.254$\pm$0.024 (0.083) & 0.243 & 0.085$\pm$0.030 ($-$0.037) & 1, 7, 8, 9\\
B1422+231     & cusp  & 77.0  & 1.643  & 0.187$\pm$0.004 (0.110) & 0.636 & $-$0.030$\pm$0.004 ($-$0.131) & 1, 10, 11, 3\\
B1555+375     & fold  & 102.6 & 1.735  & 0.417$\pm$0.026 (0.199) & 0.365 & 0.235$\pm$0.028 (0.023) & 1, 12 \\
B1608+656$^{\dagger\dagger}$     & fold  & 99.0  & 1.997  & 0.492$\pm$0.002 (0.568) & 0.831 & 0.327$\pm$0.003 (0.411) & 13, 14 \\
B1933+503$^\dagger$     & fold  & 143.0 & 1.605  & 0.389$\pm$0.017 (0.040)    & 0.884 & 0.656$\pm$0.009 (0.257) & 15, 16, 17 \\
B2045+265     & cusp  & 34.9  & 0.762  & 0.501$\pm$0.020 (0.030) & 0.253 & 0.267$\pm$0.027 ($-$0.163) & 1, 9, 18, 19 \\
%Q2237+030   & cross & 0.85 & 146.3 & 1.83 & 0.36$\pm$? (0.39) & 2, 13 \\
\hline
\end{tabular}
\\ Notes: the quoted $\Rcusp$ and $\Rfold$ values in Col. 5 and 7 are
measured at the radio wavelengths; their uncertainties are derived
from the uncertainties in flux measurements (see Table A1 for the
measured fluxes of the close triple images). Values in the parentheses
are predicted by our best-fitting lens model, see Sect. 5.1. ($\dagger$)
Flux ratios are likely affected by systematic errors due to
scattering. ($\dagger\dagger$) Quoted fluxes are after correction for
the time delays. References: (1) \citealt{Koopmans2003JVASCLASS}; (2)
\citealt{Phillips2000B0128}; (3) \citealt{FIK1999}; (4)
\citealt{Lawrence1995MG0414}; (5) \citealt{Katz1997MG0414}; (6)
\citealt{Ros2000MG0414}; (7) \citealt{Jackson98B0712}; (8)
\citealt{Jackson2000}; (9) Cfa-Arizona Space Telescope Lens Survey
(CASTLES, see http://cfa-www.harvard.edu/castles); (10)
\citealt{Impey1996B1422}; (11) \citealt{Patnaik1999B1422}; (12)
\citealt{Marlow1999B1555}; (13) \citealt{KoopmansFassnacht1999B1608};
(14) \citealt{Fassnacht1996B1608}; (15) \citealt{Cohn2001B1933}; (16)
\citealt{Sykes1998B1933}; (17) \citealt{Biggs2000B1933}; (18)
\citealt{Fassnacht1999B2045}; (19) McKean et al. 2007; .
\end{minipage}
\end{table*}

\section{Statistical flux-ratio distributions} 

In this section, we study the statistical impact of CDM substructures
on flux ratios. For this purpose, we generate mock galaxies of generic
smooth lens potentials and with morphological properties similar to
those of galaxies from SDSS. We then add to them subhalo populations
from the Aquarius (\citealt{volker08Aq}) and the Phoenix
(\citealt{Gao12Phoenix}) simulations. This enables us to forecast
$\Rcusp$ and $\Rfold$ distribution expected for a large sample of
lensed systems and study the impact of the halo properties on these
distributions. In Sect. 3.1 we present the method to model the generic
lens potentials of massive elliptical lenses, and in Sect. 3.2 how we
model their substructure populations. We describe the technique used
to mock a statistical sample of quadruply-lensed quasars in
Sect. 3.3. Finally results are given in Sect. 3.4.

\subsection{Smooth lens model}

To model the main lens halo (which is responsible for producing
quadruply-lensed images), we adopt the approach from Keeton et
al. (2003), with which we predict generic distributions for the cusp
and fold relations.

\citet{KGP2003apj,KGP2005Fold} have shown that the flux (ratio)
distributions have a weak dependence on the radial profile (from point
mass to isothermal) of the lens mass distribution, but are sensitive
to ellipticity $e$ ($\equiv1-q$, where $q$ is the axis ratio),
higher-order multipole amplitude $a_m$ and external shear $\gamma_{\rm
  ext}$. In this work, we use a generalised isothermal ellipsoidal
profile with an Einstein radius of 1.0$\arcsec$ and also take into
account the three aspects above. The detailed lens modelling and
definitions for the parameters are described in the Appendix.

For choosing $e$ and $a_m$, we use the result from \citet{HaoMao2006},
who measured ellipticities and higher-order multipoles ($m=3,~4$) of
galaxies from SDSS. The mean and scatter of these shape parameters
(mean $\bar{e}=0.23$, dispersion $\sigma_e=0.13$, mean
$\bar{a}_3=0.005$, dispersion $\sigma_{a_3}=0.004$, mean
$\bar{a}_4=0.010$, dispersion $\sigma_{a_4}=0.012$) are comparable to
the values reported for the galaxy samples used in Keeton et
al. (2003, 2005).

We note that by using the observed galaxy morphology distributions, we
implicitly assume that the shape of dark matter (and thus total)
density profiles follows baryons in the inner parts of the halo where
strong lensing occurs. This has been supported by lensing observations
from e.g., \citet{Koopmans2006apj} and \citet{Sluse2012COSMOGRAIL25}.

It is also worth noting that although we draw shape parameters ($e$
and $a_m$) from a galaxy sample at lower redshifts ($z<0.2$), as
addressed in Keeton et al. (2003, 2005), these distributions are not
expected to be significantly different from those of the observed
lensing galaxies at intermediate redshifts; observations have shown no
significant evolution in the mass assembly history of early-type
galaxies since $z\approx1$ (\citealt{Thomas2005,Koopmans2006apj}).

Finally, the lens environment (e.g., \citealt{KKS1997externalshear})
is accounted for by applying an external shear $\gamma_{\rm ext}$
drawn from a lognormal distribution with a median of 0.05 and a
dispersion of 0.2 dex, same as in Keeton et al. (2003).

When adding simulated CDM subhalos to the generalised host lens
potentials, we take 3600 different projections of subhalo
distributions (see Sect. 3.2), and add each projected distribution to
one of the host lens potentials. In order to maintain the possible
correlation between ellipticities and high-order multipoles, we draw
the combination of measured ($e$, $a_3$, $a_4$) from the observed
galaxy sample of Hao et al. (2006). For each realization, we also take
a randomly orientated external shear to add to the generalised
isothermal ellipsoid.

\subsection{CDM subhalos from the Aquarius and Phoenix simulations}

To populate smooth lens potentials with CDM substructures, we take two
sets of high-resolution cosmological $N$-body simulations: the
Aquarius (\citealt{volker08Aq}) and Phoenix (\citealt{Gao12Phoenix})
simulation suites. The former is composed of six Milky Way-sized halos
($M_{200}\sim10^{12}h^{-1}M_{\odot}$) and the latter consists of nine
galaxy cluster-sized halos ($M_{200}\sim10^{15}h^{-1}M_{\odot}$;
$M_{200}$ here is referred to as the virial mass, defined as the mass
within $R_{200}$, the radius within which the mean mass density of the
halo is 200 times the critical density of the Universe).

Observed lenses typically have an inner velocity dispersion of
$200-300$ km/s (e.g., \citealt{KT02_VDMG2016,VandeVen2003_FPVD}), and
some of them are also shown to live in the group environment (e.g.,
\citealt{Momcheva2006LensEnv,Wong2011ShearEnv}). In comparison, the
Aquarius halos have an equivalent inner velocity dispersion (estimated
by $1/\sqrt{2}$ of the peak velocity) of $\sim150$ km/s, and $\sim900$
km/s for the Phoenix halos. We rescale all fifteen halos from both
simulation suites to host halos of masses fixed at
$M_{200}=10^{12}h^{-1}M_{\odot}, ~10^{13}h^{-1}M_{\odot},~{\rm and}~
5\times10^{13}h^{-1}M_{\odot}$. By doing so, we can study the lensing
effects from subhalo populations hosted by halos on different mass
scales and their dependences on host halo properties.

To be precise, we take both simulations at their second resolution
levels, at which the minimum resolved subhalos have masses about seven
orders of magnitude below the virial masses of their hosts. We define
a rescaling factor $\mathfrak{R}$, which is the ratio between
$M_{200}$ of the arbitrary halo and that of a simulated halo. We
rescale the masses of subhalos accordingly by a factor of
$\mathfrak{R}$, and their velocities, sizes and halo-centric distances
by a factor of $\mathfrak{R}^{\frac{1}{3}}$, so that the
characteristic densities remain the same. It is noteworthy to mention
that we only use $M_{200}$ of individual halos to work out their
rescaling factor. It is the subhalos (not the main halos) that we
rescale and add to the constructed host lens potentials (as described
in Sect. 3.1).

In the following, we present the rescaled subhalo properties,
including mass function, characteristic velocities, sizes and spatial
distributions.

\subsubsection{Subhalo mass function}

From Sect. 4.1 and Fig. 13 of Gao et al. (2012), no significant
difference is seen between the shapes of subhalo mass functions of
cluster-sized Phoenix halos and of Milky Way-sized Aquarius halos.
The number of Phoenix subhalos is higher by 35\% than the number of
Aquarius subhalos at any fixed subhalo-to-halo mass ratio $m_{\rm
sub}/M_{200}$. This is because clusters are dynamically younger than
galaxies, therefore there are more subhalos surviving the tidal
destruction.

\subsubsection{Spatial distributions and projection effects}

From Sect. 4.2 and Fig. 15 of Gao et al. (2012), the spatial
distribution of the Phoenix subhalos is slightly more concentrated
(more abundant near the centre) than that of the Aquarius subhalos due
to the assembly bias effect, as the Phoenix simulations start from
high density regions.

For this work, the projected spatial distribution of subhalos,
especially in terms of the radial distribution of their surface number
density, is of particular interest, as it directly influences the total
lensing cross-section from subhalos. In this subsection, we show the
mean projected spatial distributions obtained from averaging over
hundreds of projections per host halo from both simulation suites.

\begin{figure} 
\centering  
\includegraphics[width=8cm]{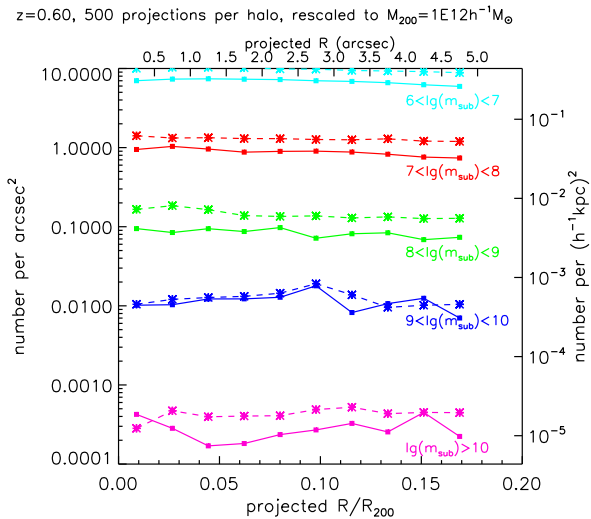} 
\caption{The radial distributions of projected subhalo number
  densities, averaged over six Aquarius halos (solid lines) and nine
  Phoenix halos (dashed lines) at redshift $z=0.6$; 500 random
  projections are used per halo. All subhalo populations are rescaled
  to $10^{12} h^{-1}M_{\odot}$. Five different subhalo-mass ranges
  have been inspected: $10^{6\sim7}h^{-1}M_{\odot}$ (cyan),
  $10^{7\sim8} h^{-1}M_{\odot}$ (red), $10^{8\sim9} h^{-1}M_{\odot}$
  (green), $10^{9\sim10} h^{-1}M_{\odot}$ (blue) and $>10^{10}
  h^{-1}M_{\odot}$ (pink). The axis at the top gives the projected
  radius in arcsec; the one at the bottom gives the projected radius
  normalized to $R_{200}$. The axis on the left gives number per
  sq. arcsec; the one on the right gives number per sq. $h^{-1}$kpc
  (in physical scale).
}\label{fig:SubProjDistribution} 
\end{figure}

Fig. \ref{fig:SubProjDistribution} shows the projected subhalo number
densities as a function of (projected) halo-centric distance up to
0.2$R_{200}$. An important feature of the distribution is that it
varies little with the projected halo-centric distances. Note that
this is true (only) at smaller radii from the host centre and is also
a result of the projection effect. More massive subhalos, e.g.,
$m_{\rm sub}\ga10^9h^{-1}M_{\odot}$, can only survive in the outer
region of their host halo because of tidal destruction; their presence
within the projected central $\sim0.1R_{200}$ is purely due to chance
alignment. We refer the reader to Springel et al. (2008, Fig. 11 and
discussion therein) for the 3$D$ spatial distribution of the subhalo
population.

Also can be seen from Fig. \ref{fig:SubProjDistribution} is that as
the subhalo mass decreases by one decade, there is an increase by
roughly a factor of ten in the number density of (projected) subhalos,
i.e., $dN/d\ln m_{\rm sub}\propto m_{\rm sub}^{-1}$. This is in fact
expected from the subhalo mass function ($dN/dm_{\rm sub}\propto
m^{-1.9}$, Springel et al. 2008), where the logarithmic slope is close
to $-$2.0. In Fig. \ref{fig:SubProjDistribution}, the Aquarius and
Phoenix subhalos are rescaled to a host mass of
$10^{12}h^{-1}M_{\odot}$; but the same features are also seen when
they are rescaled to a host mass of $\ga10^{13}h^{-1}M_{\odot}$.

%There are two major features in the plot: first, the distribution is
%very flat at smaller projected halo-centric distance; second as the
%subhalo mass decreases by one decade, there is an increase by roughly
%a factor of ten in the number density of (projected) subhalos, i.e.,
%$dN/d\ln m_{\rm sub}\propto m_{\rm sub}^{-1}$. This is in fact
%expected from the subhalo mass function ($dN/dm_{\rm sub}\propto
%m^{-1.9}$, Springel et al. 2008), where the logarithmic slope is close
%to $-$2.0. In Fig. \ref{fig:SubProjDistribution}, the Aquarius and
%Phoenix subhalos are rescaled to a host mass of
%$10^{12}h^{-1}M_{\odot}$; but the same features are also seen when
%they are rescaled to a host mass of $\ga10^{13}h^{-1}M_{\odot}$.

\begin{figure*}
\centering
\includegraphics[width=8cm]{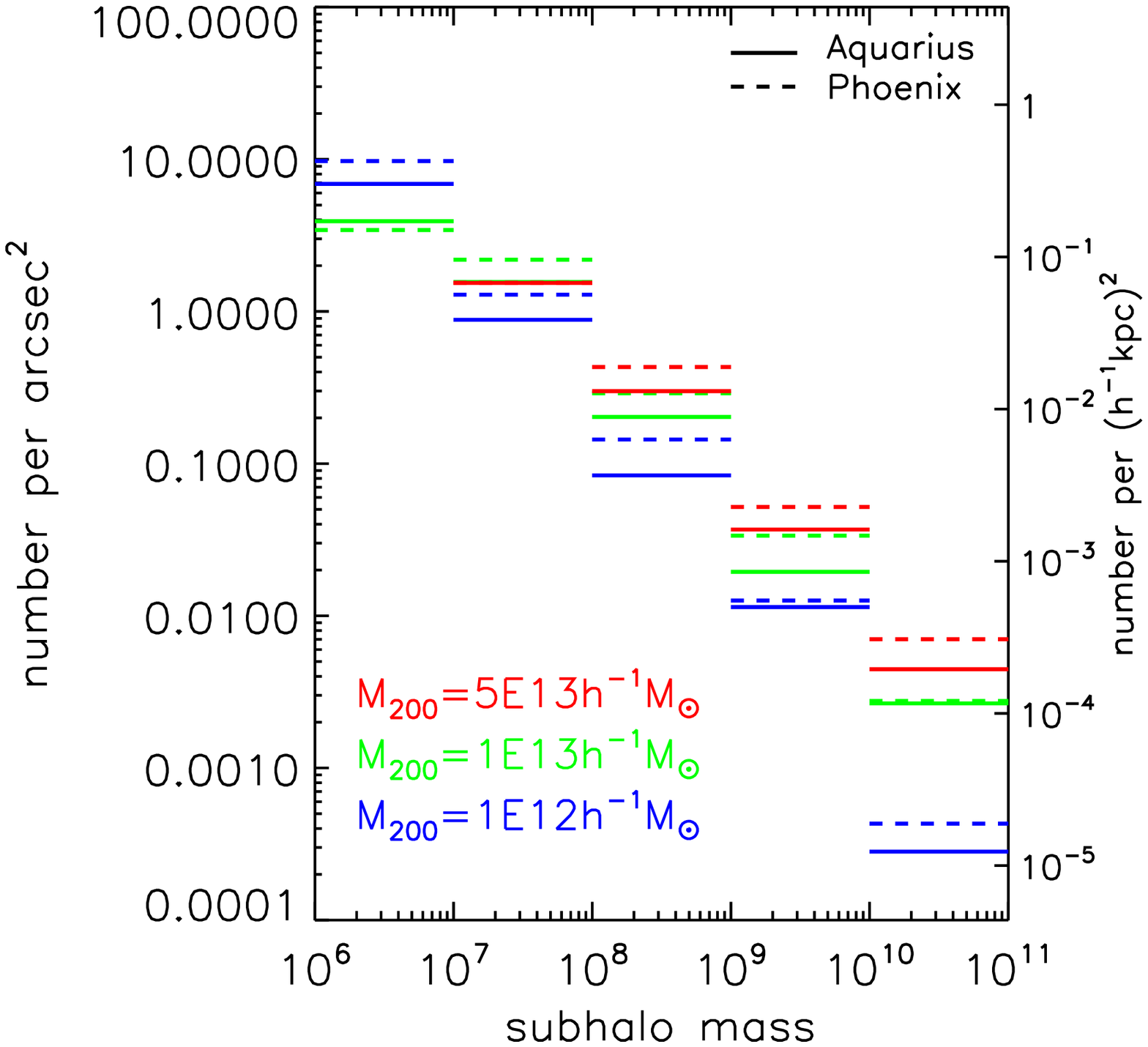}
\includegraphics[width=8cm]{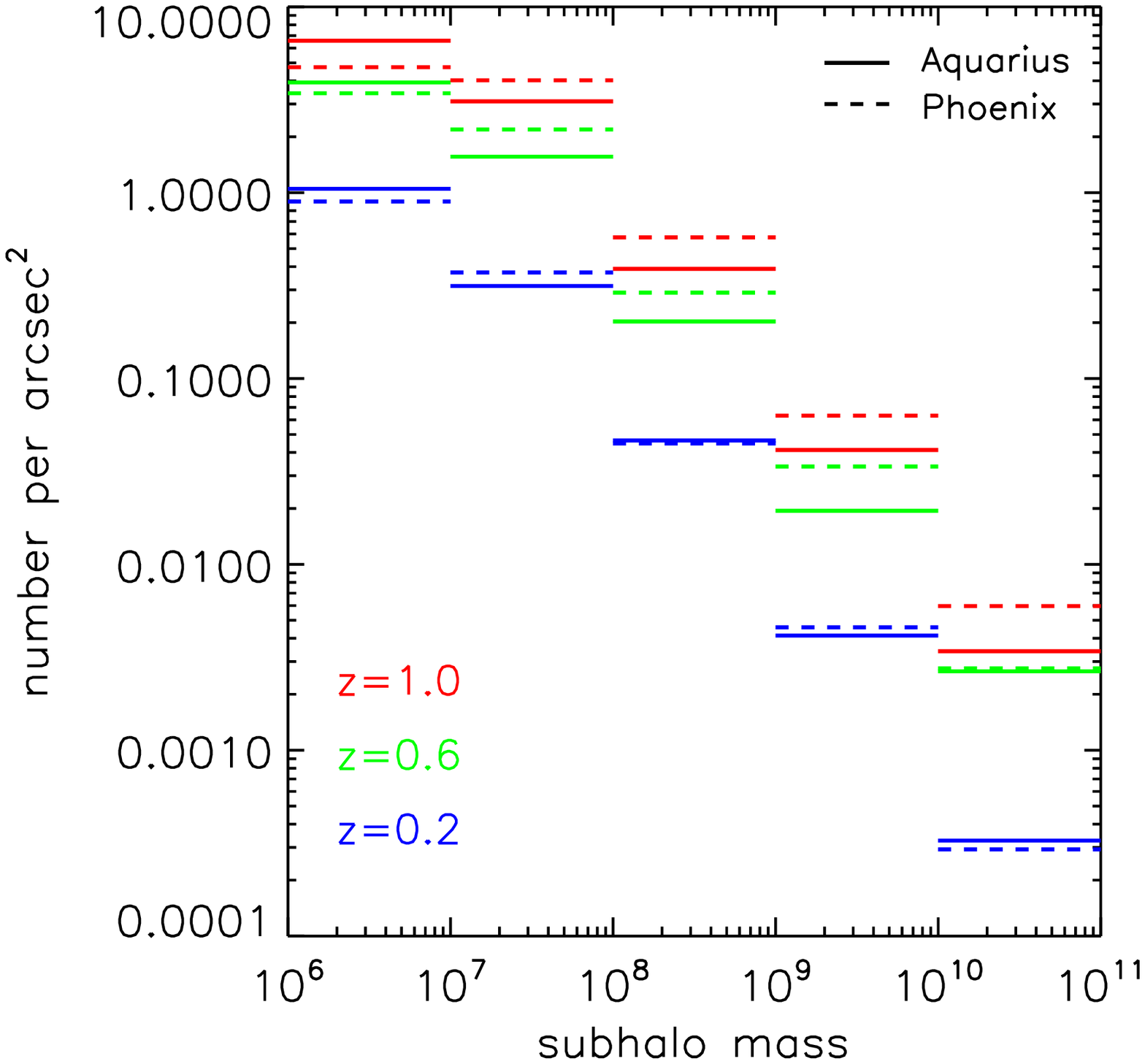}
\caption{Projected subhalo number densities averaged within the
  central $R\leqslant 5\arcsec$ region, as a function of subhalo
  masses. The panel on the left shows the host mass dependence:
  subhalos taken at $z=0.6$, their hosts rescaled to $M_{200}=10^{12}
  h^{-1}M_{\odot}$ (blue), $10^{13} h^{-1}M_{\odot}$ (green) and
  $5\times 10^{13} h^{-1}M_{\odot}$ (red). The right-hand side panel
  shows the redshift dependence: host halos are rescaled to
  $M_{200}=10^{13} h^{-1}M_{\odot}$, taken at $z=0.2$ (blue), $z=0.6$
  (green) and $z=1.0$ (red). For both panels, 500 random projections
  are used per halo. The axis on the left-hand side of each panel
  gives number per sq. arcsec. The axis on the right-hand side of the
  left panel also gives number per sq.  $h^{-1}$kpc (in physical scale
  corresponding to a redshift at $z=0.6$). Solid lines show the number
  densities of the Aquarius subhalos; dashed lines are for the Phoenix
  subhalos.}
  \label{fig:SubSpatialMassRed}
\end{figure*}
As the projected subhalo number densities remain constant in the inner
part of a host halo, we take the mean values averaged within the
central $R\leqslant5\arcsec$ region and studied their dependences on
host halo mass and redshifts. Fig. \ref{fig:SubSpatialMassRed} shows
such mean number densities as a function of subhalo mass, plotted for
host halos at three different $M_{200}$ and three different
redshifts. It can be seen that rescaling to more massive host halos
will result in a higher number density of projected subhalos; the
number per sq. arcsec also increases significantly with redshift.

Note that in Fig. \ref{fig:SubSpatialMassRed} the lowest-mass bins
below $10^{7}h^{-1}M_{\odot}$ are only complete for host halos of
$M_{200}=10^{12}h^{-1}M_{\odot}$. Due to the simulation resolution
limit, a level-two halo rescaled to $M_{200}=10^{12}h^{-1}M_{\odot}$,
$10^{13}h^{-1}M_{\odot}$ and $5\times10^{13}h^{-1}M_{\odot}$ would
only host a complete subhalo sample down to a mass of
$\sim2\times10^{5}h^{-1}M_{\odot}$, $\sim2\times10^{6}h^{-1}M_{\odot}$
and $\sim10^{7}h^{-1}M_{\odot}$, respectively. Above these
``completeness'' mass scales, one can easily read off the projected
number densities $\eta_{\rm sub}$ for group-sized host halos
($M_{200}\ga10^{13}h^{-1}M_{\odot}$), which satisfy:
\begin{equation}
\frac{d\eta_{\rm sub}}{d\ln m_{\rm sub}}\approx0.01~\left(\frac{m_{\rm
    sub}}{3\times10^{8}h^{-1}M_{\odot}}\right)^{-1}(h^{-1}\kpc)^{-2}.
\end{equation} 
The surface mass density in each mass decade is then estimated to be
$\approx3\times10^{6}h^{-1}M_{\odot}(h^{-1}\kpc)^{-2}$. Consider a
typical lens system with lens and source redshifts $z_{\rm l}=0.6$ and
$z_{\rm s}=2.0$, the critical surface mass density is $\Sigma_{\rm
  cr}\approx 3\times10^9h^{-1}M_{\odot}(h^{-1}\kpc)^{-2}$; then the
surface mass fraction in substructures over five mass decades above
$10^6h^{-1}M_{\odot}$ amounts to $\la1\%$ around the critical curve,
where the local convergence is $\kappa_{\rm
  cr}\equiv\Sigma/\Sigma_{\rm cr}$ $\approx$ 0.5. We mention in
passing that the different subhalo mass fraction between here and
0.3\% as in Xu et al. (2009) is attributed to a richer subhalo
populations of group-sized halos considered here. These fractions are
also consistent with \citet{Vegetti2014SubFrac}, who searched for
imprints of substructures in arc images of 11 gravitational lens
systems from the Sloan Lens ACS Survey.

\subsubsection{Subhalo density profiles}

The peak circular velocity $V_{\rm max}$ and the radius $r_{\rm
max}$, at which $V_{\rm max}$ is reached, are two important shape
parameters for a subhalo's density profile. As can be seen from
Fig. 14 of Gao et al. (2012), the relation between $V_{\rm max}$ and
$r_{\rm max}$ is the same for the Aquarius and the Phoenix subhalos.

Springel et al. (2008) studied the density profile of subhalos and
found them to be well fitted by Einasto profiles (\citealt{Einasto65})
with slope parameter $\alpha=0.18$,
\begin{equation}
\rho(r)=\rho_{-2} \exp
\left(-\frac{2}{\alpha}\left[\left(\frac{r}{r_{-2}}\right)^{\alpha}-1\right]\right),
\end{equation}
where $\rho_{-2}$ and $r_{-2}$ are the density and radius at which the
local slope is $-$2. For $\alpha=0.18$, $\rho_{-2}$ and $r_{-2}$ are
related to $V_{\rm max}$ and $r_{\rm max}$ by $r_{\rm
  max}=2.189~r_{-2}$ and $V^2_{\rm max}=11.19~G r^2_{-2}\rho_{-2}$,
where $G$ is the gravitational constant (see e.g., Springel et
al. 2008 for more details about fitting Einasto profiles).

From both simulation sets, instead of taking particle distributions of
subhalos for ray tracing, we take the measured $V_{\rm max}$ and
$r_{\rm max}$ for each subhalo and assume an Einasto profile with
$\alpha=0.18$. We truncate the profile at a truncation radius $r_{\rm
  trnc}$, which is set to be two times the half-mass radius $r_{\rm
  half}$ of the subhalo ($r_{\rm trnc}=2r_{\rm half}$); the mass
enclosed within such a truncation radius differs from the quoted
subhalo mass $m_{\rm sub}$ by less than 10\%. For subhalos below the
resolution limit, we present the detailed method of deriving their
profile parameters in Sect. 4.

We note that the Einasto parameter $\alpha$ could vary for different
subhaloes (\citealt{Vera-Ciro2013}), and that $r_{\rm max}$ cannot be
measured as accurately as $V_{\rm max}$, especially for lower-mass
subhalos. To see any potential change in the final result due to
inaccurate measurements of subhalo profiles, we simply set $r_{\rm
  max}$ of each subhalo to be 0.5, 1 and 2 times its current value and
carry out the same lensing calculations.

Here we verify that there is no significant quantitative difference in
the final flux ratio probability distributions resulting from
different adoptions of $r_{\rm max}$. But we caution that when
fundamentally different density profiles (in an extreme case a point
mass) are chosen, the violation probabilities strongly depend on
subhalo mass concentration (e.g.,
\citealt{RozoZentner2006subhaloMagPert, Chen2011CuspViolation,
  Xu2012LOS}), which is not further discussed in this paper.

\subsection{Generating a statistical sample of quadruply-lensed quasars}

In this section, we predict the statistical distribution of the flux
ratios ($\Rcusp$ and $\Rfold$) of the quadruple images of background
quasars. To this end, we mock a large sample of quadruply-lensed
quasars assuming that they are point sources, which induce more
violations to the cusp and fold relations than finite-sized sources.

For all the calculations presented in this section, we only take the
(rescaled) Aquarius and Phoenix subhalos above
$10^{7}h^{-1}M_{\odot}$. For each subhalo that has a mass $m_{\rm
  sub}^{78}\in [10^{7}h^{-1}M_{\odot},~ 10^{8}h^{-1}M_{\odot}]$ and is
projected in the central strong lensing region, we artificially
generate another ten subhalos each with a mass of $0.1\times m_{\rm
  sub}^{78}$, projected at the same halo-centric distance but with a
random azimuthal angle. By doing so, we include a complete sample of
subhalos at the $10^{6}h^{-1}M_{\odot}$ scale.

There are two reasons for this choice of the lower mass limit. First,
based on some simple finite source-size argument (e.g.,
\citealt{Xu2012LOS}) such a mass scale was commonly used as subhalo
lower mass limit in previous studies; the same adoption here will
allow us to directly compare our results to those studies. Second, due
to the nature of the subhalo mass function, the calculation done for
low-mass subhalos will be significantly more expensive than that of
their higher-mass counterparts. Therefore we neglect the contribution
from subhalos below $10^{6}h^{-1}M_{\odot}$ for the general
statistical calculations here, but in Sect. 4, we carry out case
studies using specific lens models to investigate the lensing effects
from subhalos at several mass decades below $10^{6}h^{-1}M_{\odot}$.

To eliminate biases due to halo-to-halo variations, we take a total of
3600 different projections of the simulated and rescaled subhalo
distributions (over all redshifts) and add them to the generalised
host lens potentials. To be precise, 300 projections were used for
each of the six Aquarius halos and 200 projections for each of the
nine Phoenix halos at different redshifts.

We assume the quasar redshift to be $z_{\rm s}=2.0$ and take simulated
subhalo populations at five different lens redshifts: $z_{\rm
  l}=[0.2,~0.4,~0.6,~0.8,~1.0]$, which follows the lens redshift span
of the CLASS survey. We test two different redshift distributions: (1)
a flat redshift distribution for the simulated subhalo populations,
i.e., 60/40 projections per Aquarius/Phoenix halo at each of the five
redshifts; and (2) a lensing cross-section weighted redshift
distribution assuming the main lens to be a singular isothermal sphere
with velocity dispersion $\sigma_{\rm SIS}=300$ km/s (and $z_{\rm
  s}=2.0$), which results in [26, 63, 79, 74, 58] projections per
Aquarius halo and [17, 42, 53, 50, 38] projections per Phoenix halos
at the five fixed redshifts, respectively.  In terms of the final
flux-ratio distributions, there is no significant difference between
these two lens redshift distributions. In the subsections below we
therefore present results obtained using the redshift distribution
weighted by lensing cross-sections.

For each constructed lens potential, we carry out standard lensing
calculations, similar to those used in our previous studies: a grid
with resolution of 0.005$\arcsec$/pixel covers the lens plane, where
deflection angles and second-order derivatives of the lens potentials
from the host lens and from subhalos are calculated and tabulated on
to the mesh. The adopted lens-plane resolution guarantees that the
surface density distribution of the least massive subhalos at $m_{\rm
  sub}\sim10^{6}h^{-1}M_{\odot}$ are resolved by a few to ten pixels
at radii where their half masses and peak deflection angles are
reached (see Fig. \ref{fig:EinastoProfileLowMassRes} in Sect. 4).

Source positions are uniformly distributed inside the tangential
caustic (of each constructed lens) with a number density of 20000 per
sq. arcsec, which naturally ensure that each realization is weighted
by its four-image cross section in the source plane. The lensed images
for any point source are found by solving the lens equation with a
Newton-Raphson iteration method, setting the convergence error on
image positions to be $0.0001\arcsec$. We do not consider
magnification bias in our statistical analysis; possible consequences
are discussed in a later section. In total we generate
$\sim5\times10^6$ four-image lens systems for final inspection of the
cusp and fold violations.

\subsection{Overall flux-ratio probability distributions}

%%%To here DD!
\begin{figure*}
\centering
\includegraphics[width=5.5cm]{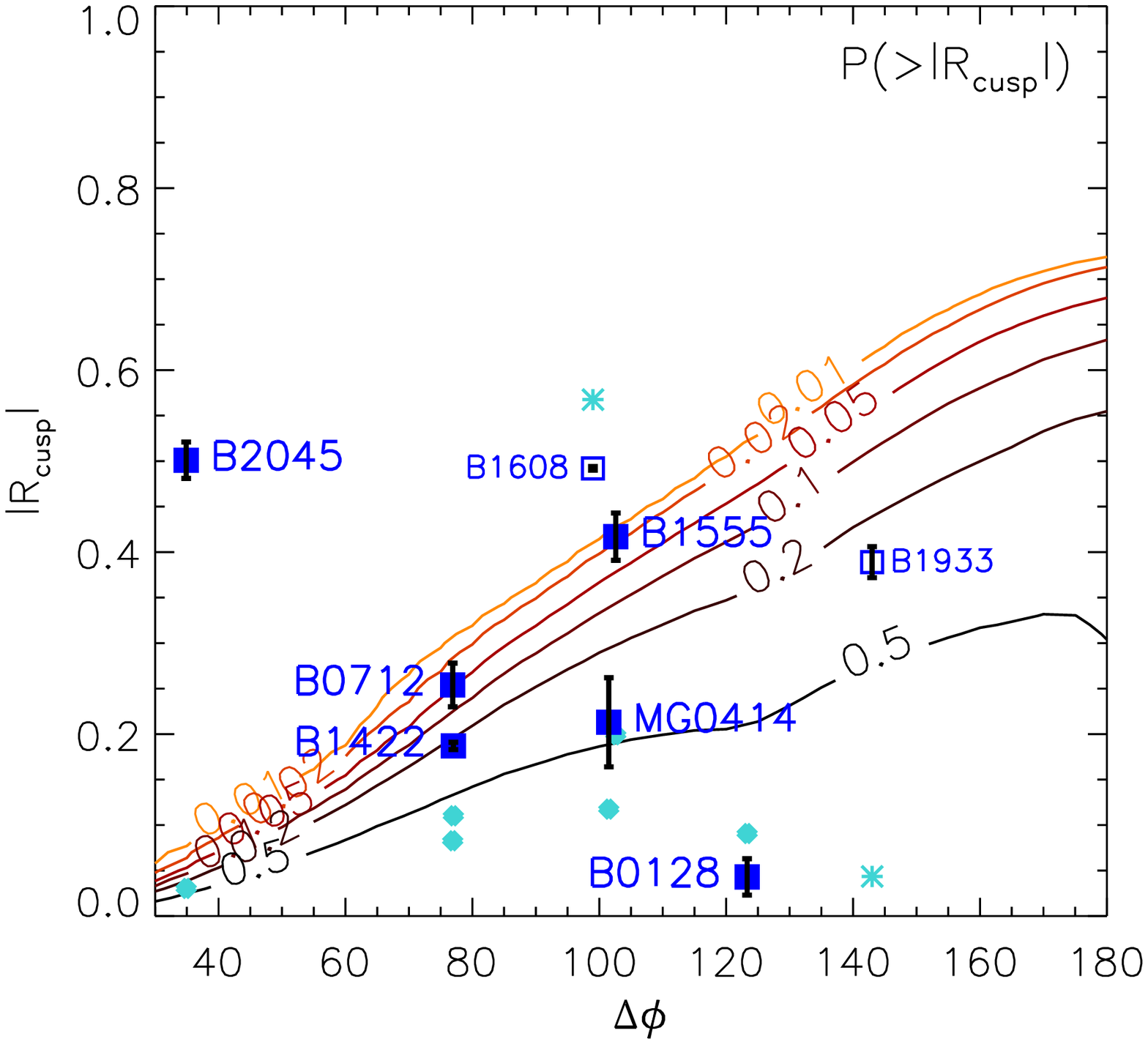}
\includegraphics[width=5.5cm]{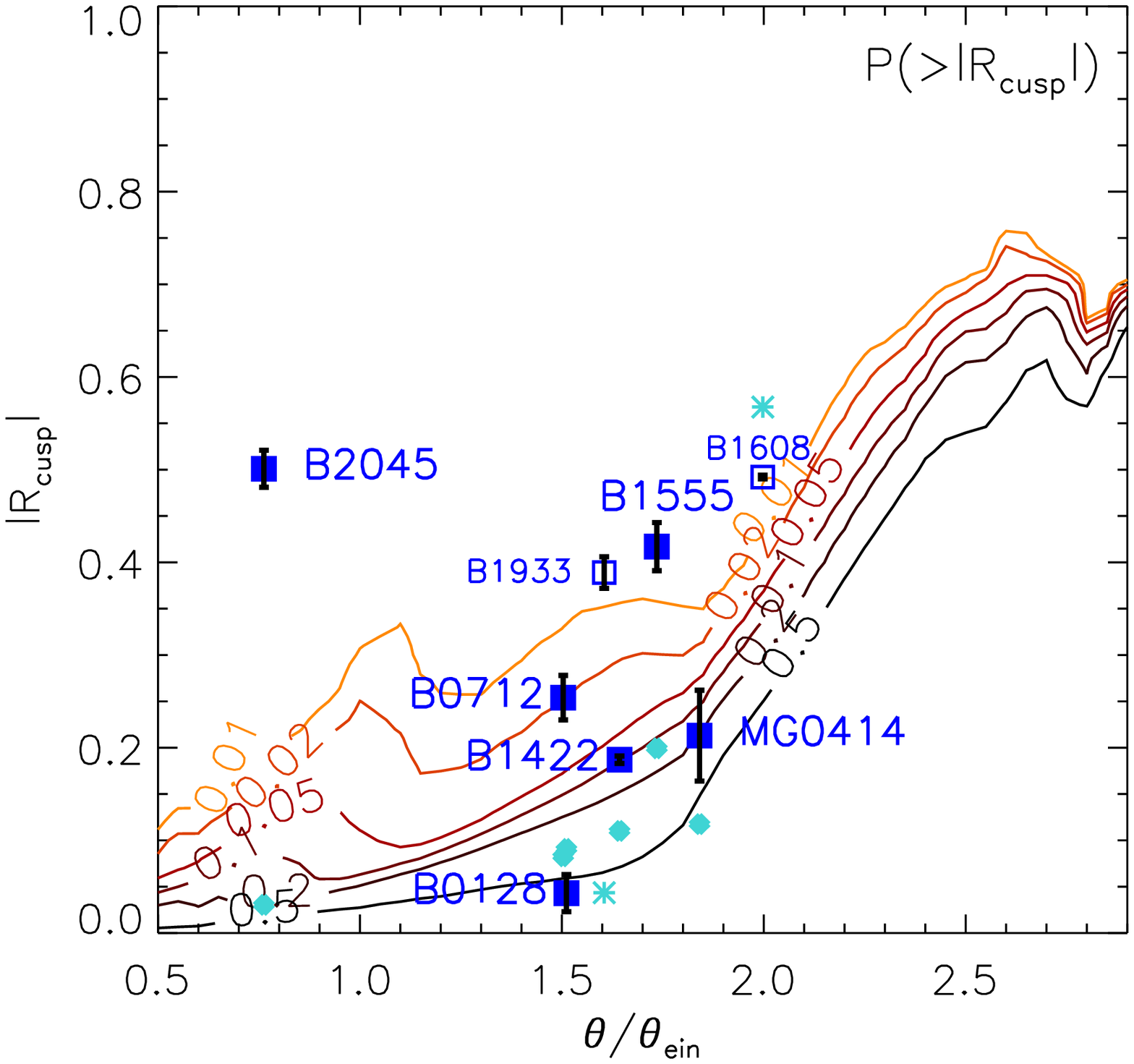}
\includegraphics[width=5.5cm]{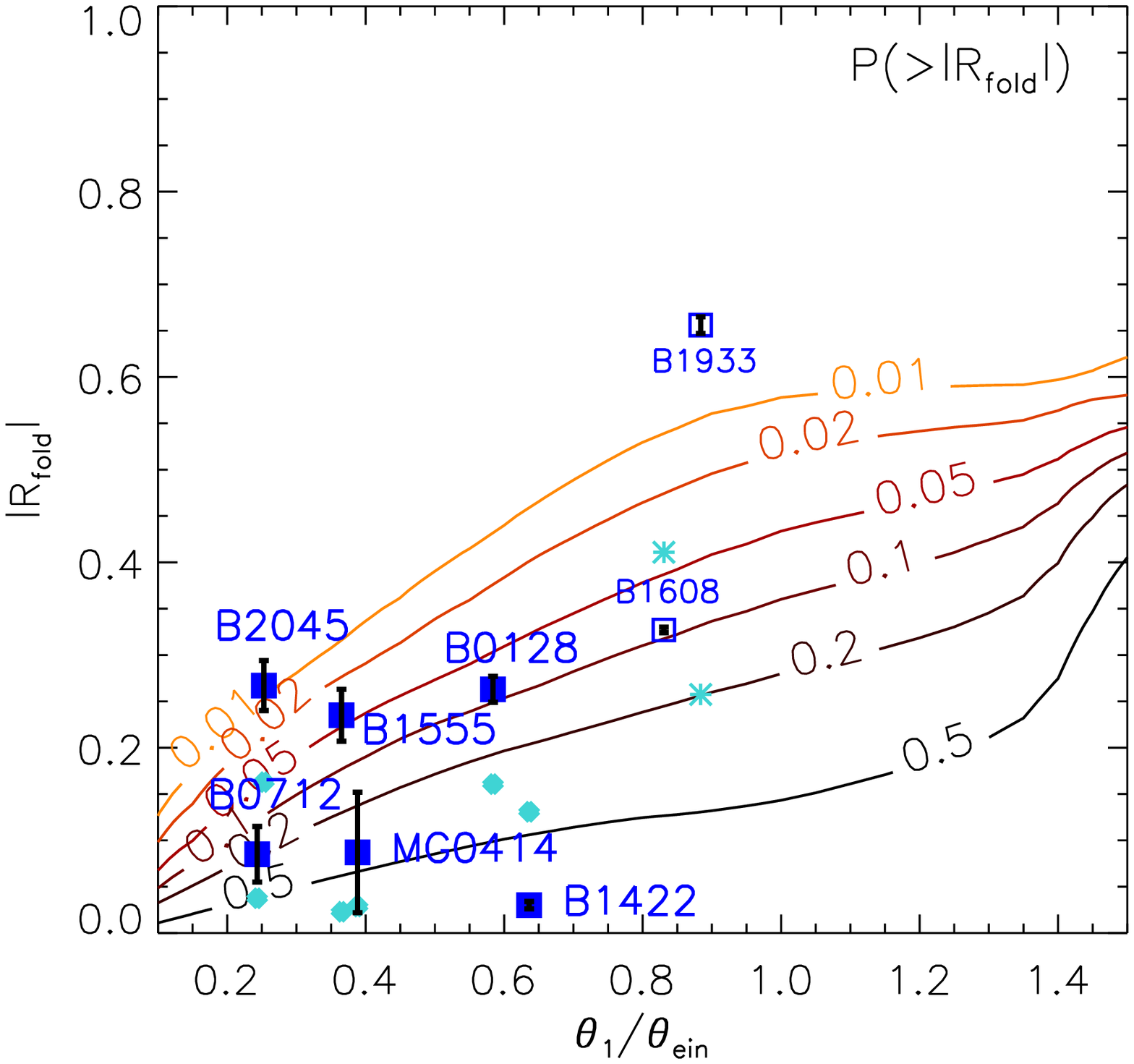}\\
\includegraphics[width=5.5cm]{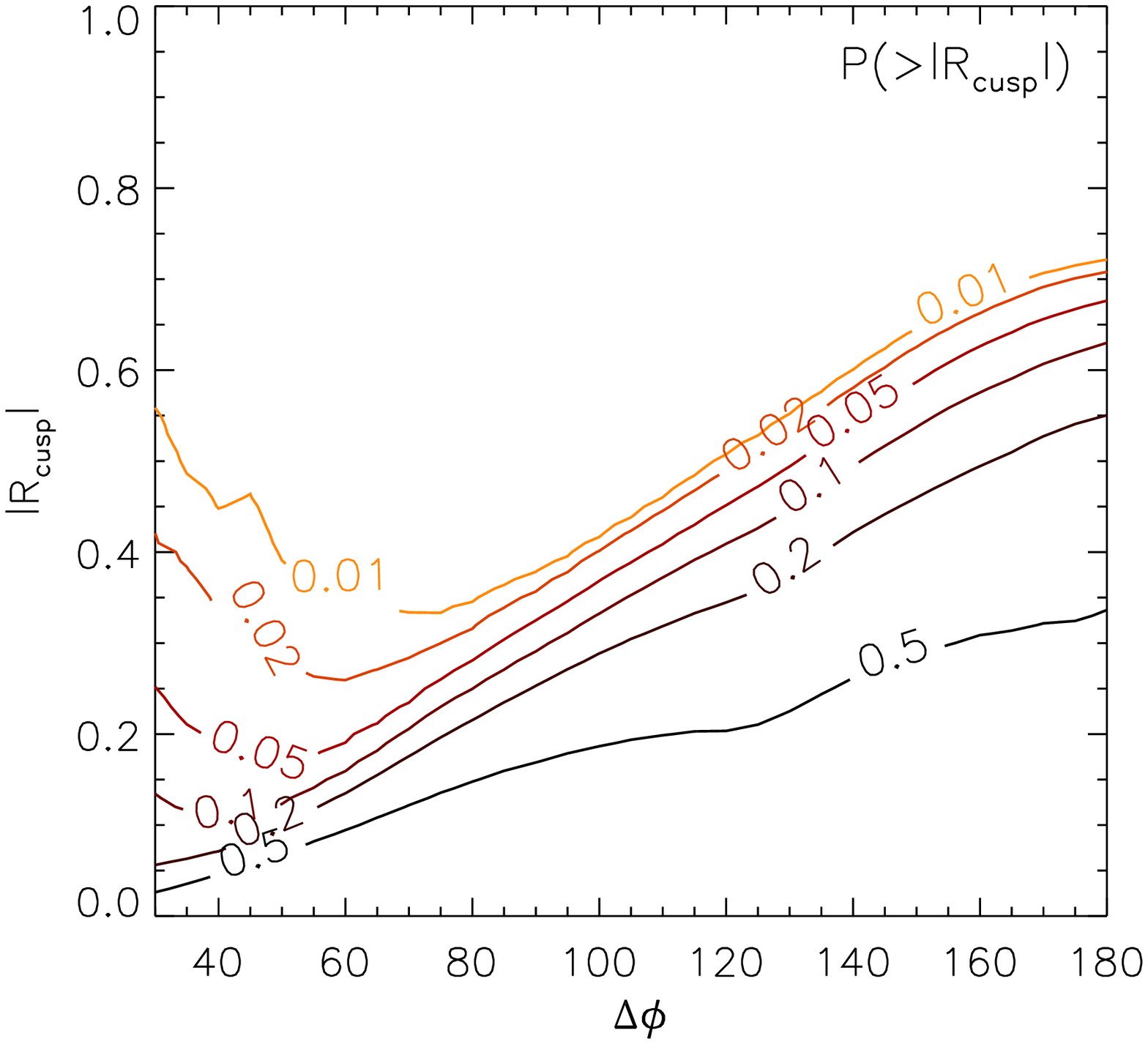}
\includegraphics[width=5.5cm]{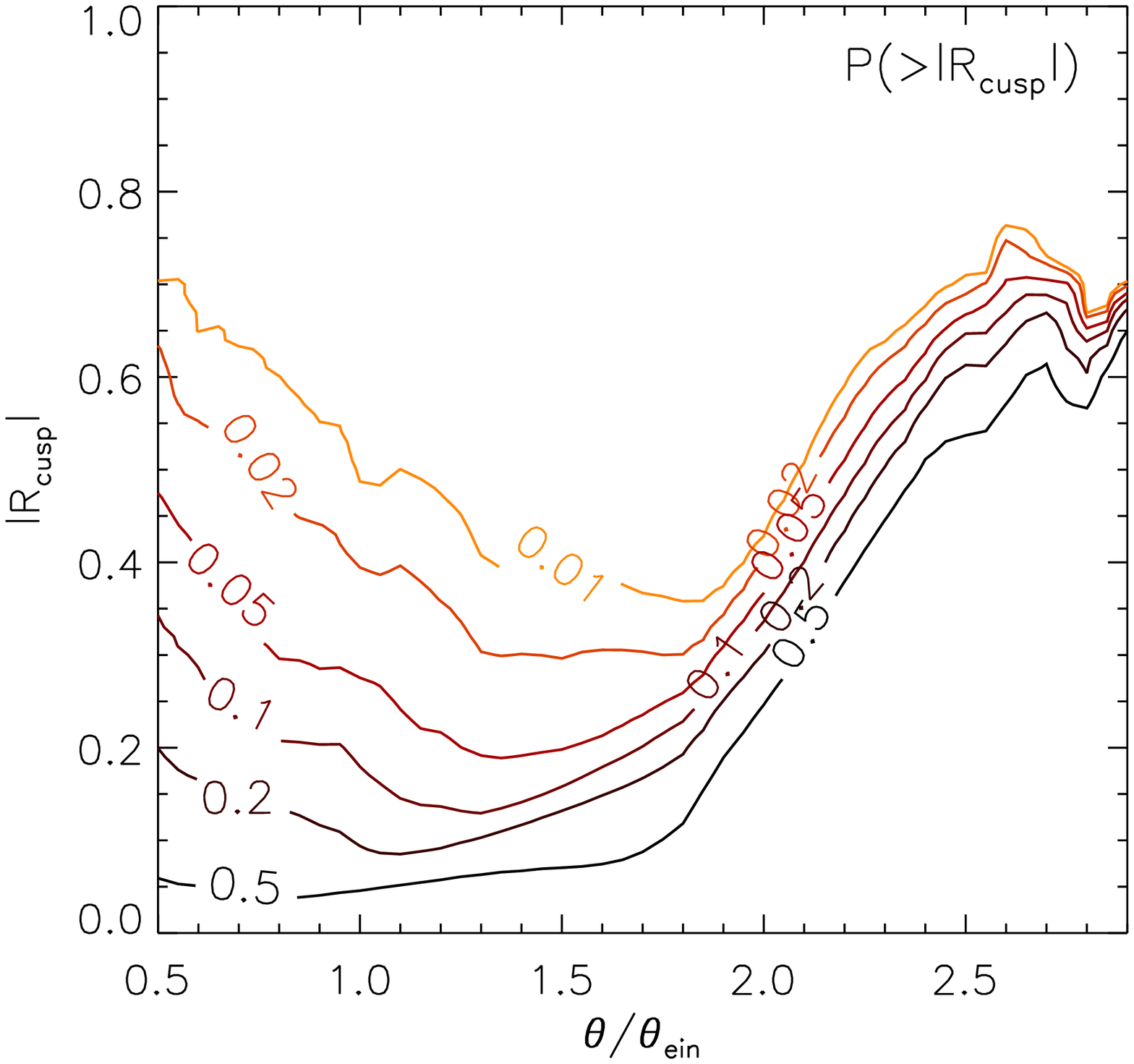}
\includegraphics[width=5.5cm]{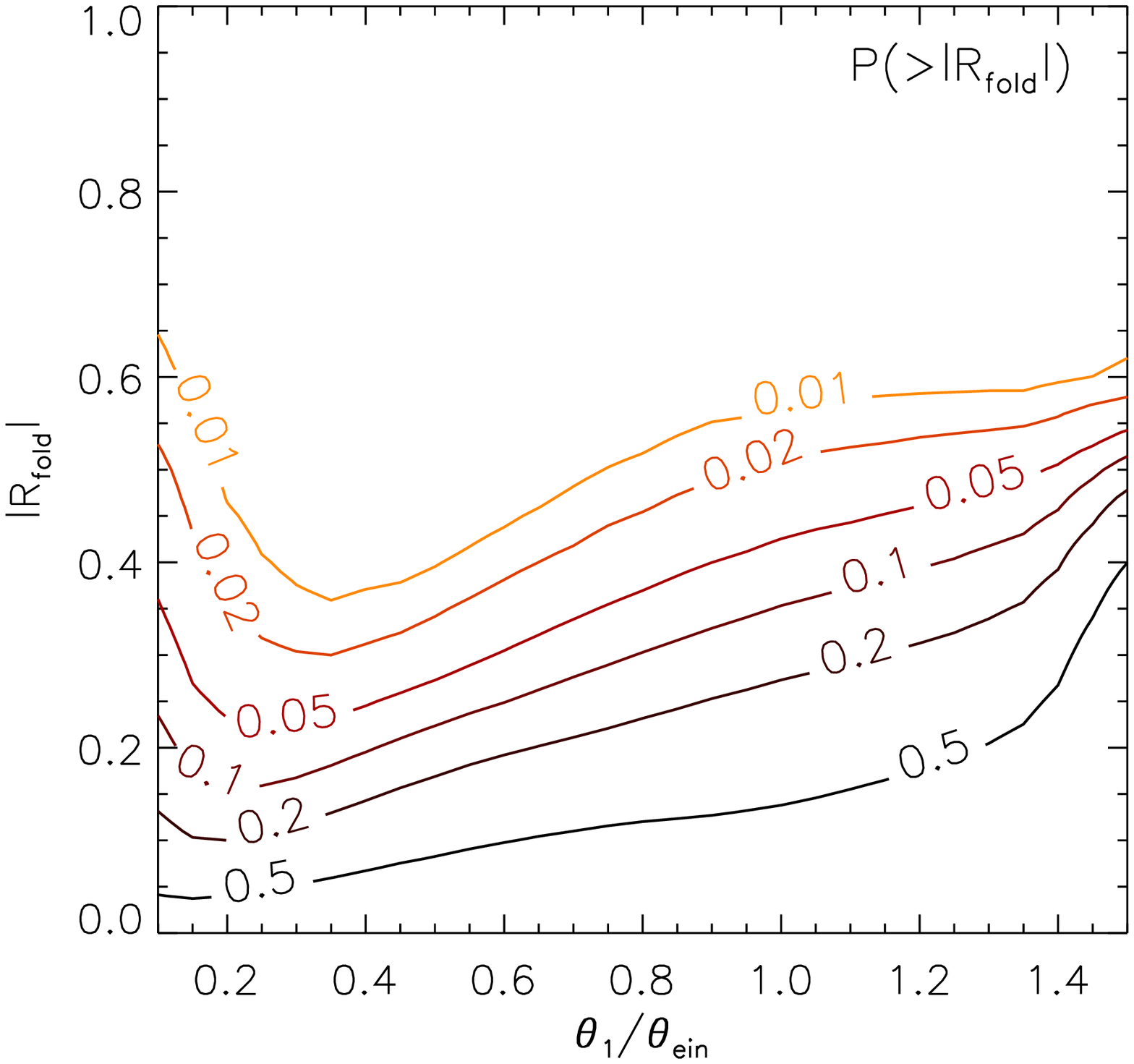}\\
\includegraphics[width=5.5cm]{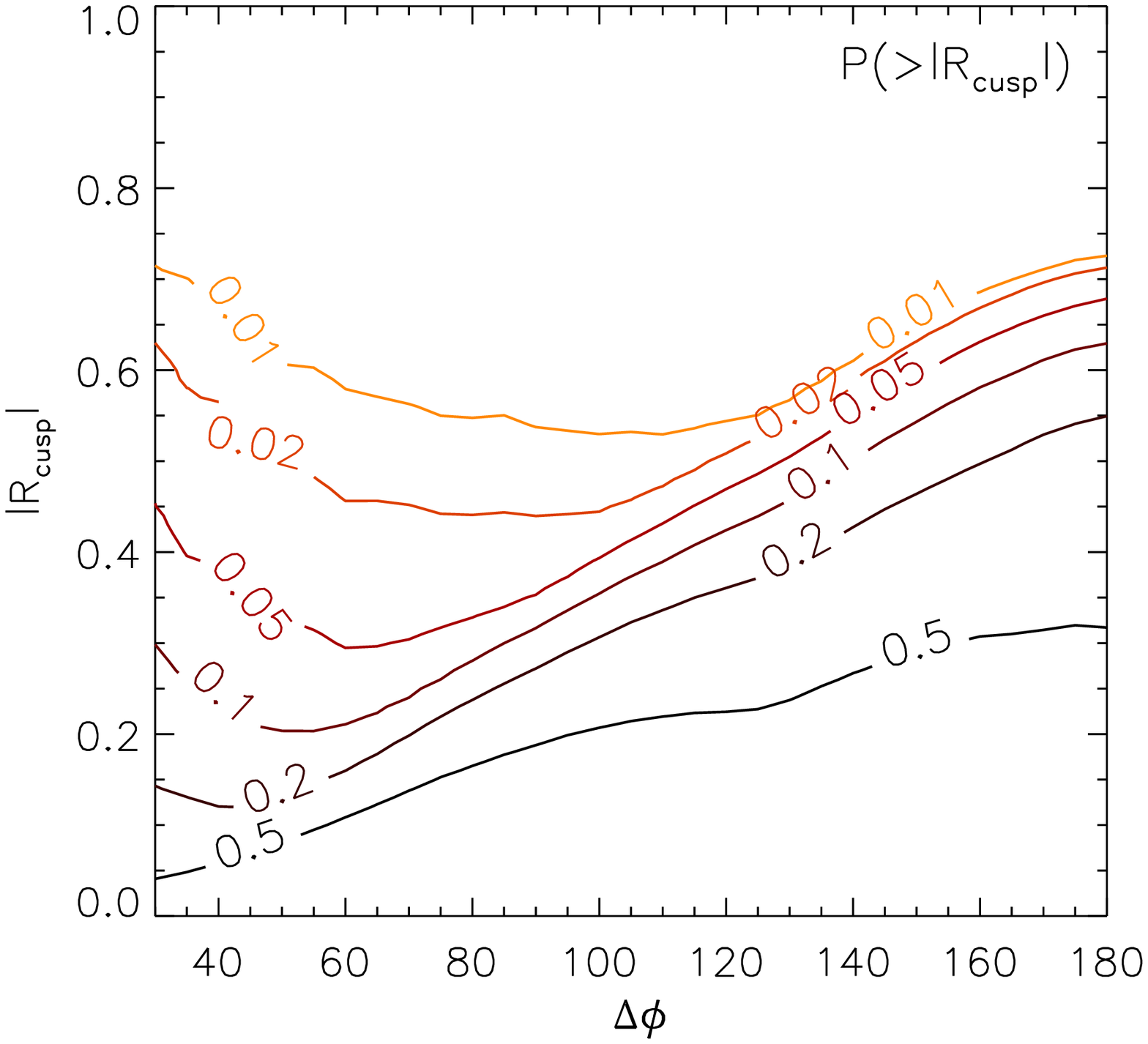}
\includegraphics[width=5.5cm]{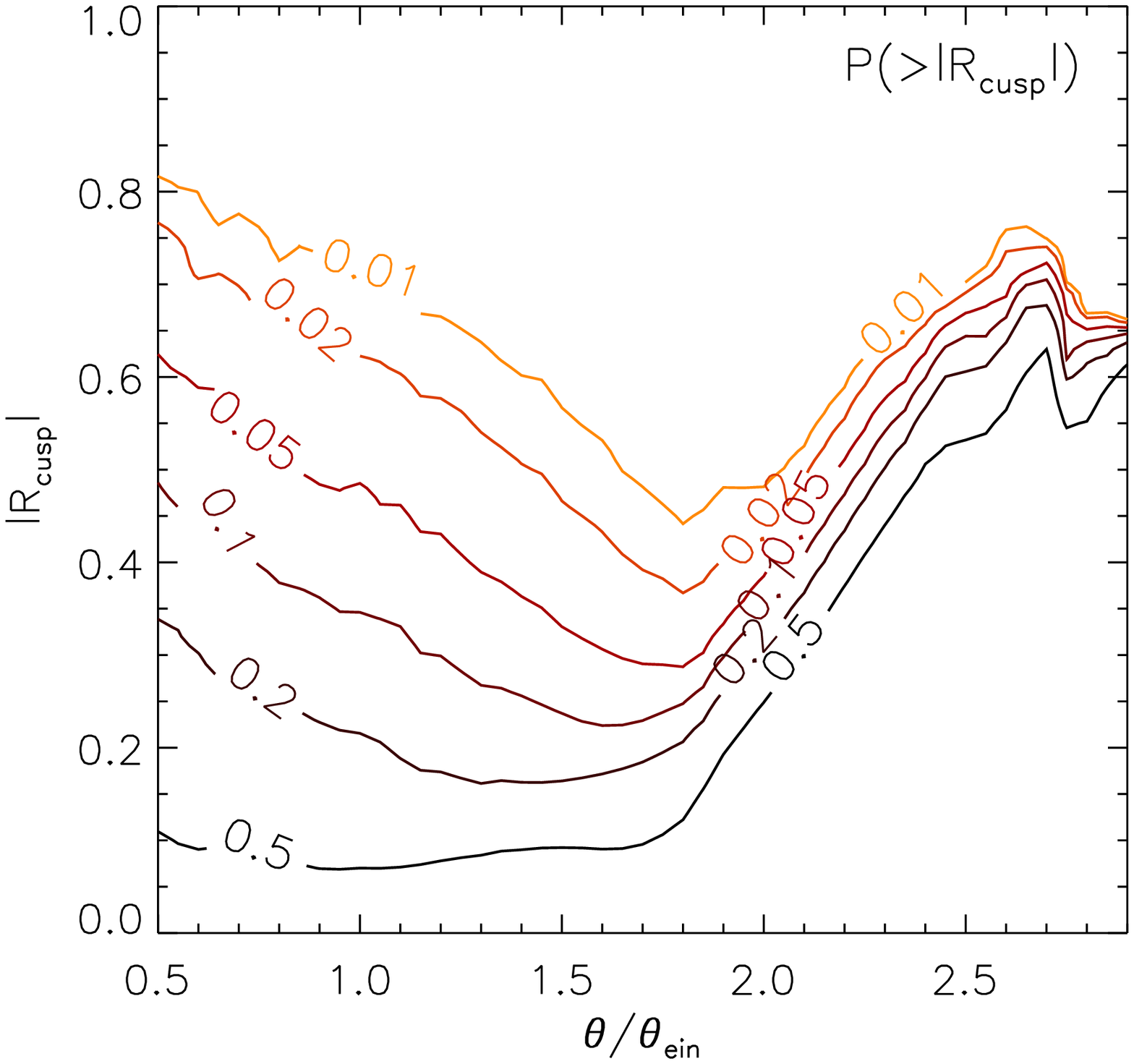}
\includegraphics[width=5.5cm]{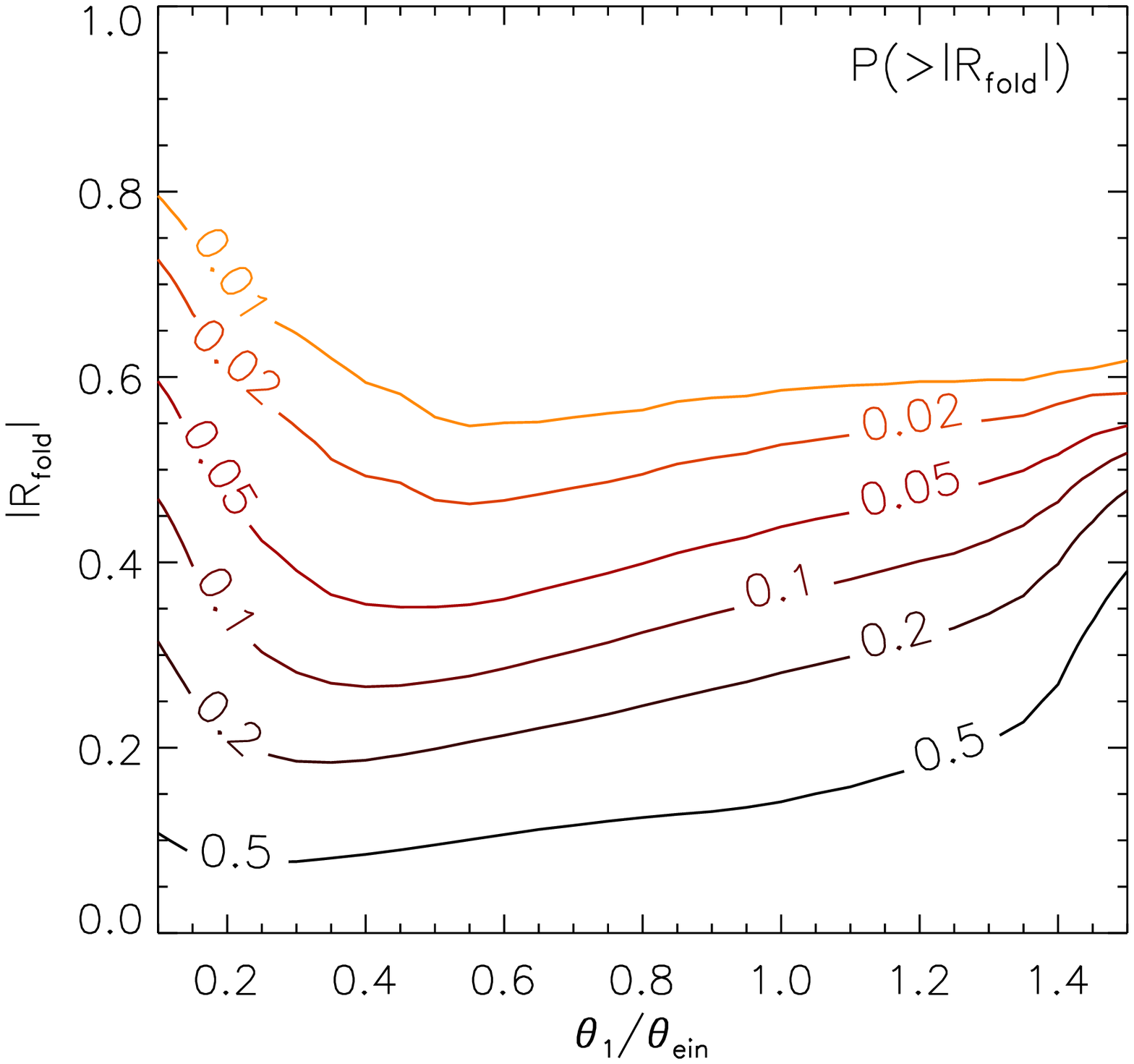}\\
\caption{Probability contour maps of conditional probabilities
  $P(>|\Rcusp|)$ for given $\Delta\phi$ (left column), $P(>|\Rcusp|)$
  for given $\theta/\Rein$ (middle column) and $P(>|\Rfold|)$ for
  given $\theta_1/\Rein$ (right column). The meanings for
  $\Delta\phi$, $\theta/\Rein$ and $\theta_1/\Rein$ can be seen in
  Fig. 1. Contour levels of 1, 2, 5, 10, 20 and 50 per cent (from
  light to dark) are plotted. Top: singular isothermal ellipsoidal
  potentials with axis ratio $q$ and higher-order perturbation
  amplitudes $a_m$ drawn from 847 observed galaxies (Hao et al. 2006),
  plus randomly oriented external shear.  Middle: smooth potentials
  (as above) plus perturbations from a simulated subhalo population
  hosted by a Milky Way-sized halo of
  $M_{200}=10^{12}h^{-1}M_{\odot}$. Bottom: smooth potentials (as for
  the top panel) plus perturbations from a simulated subhalo
  population hosted by a group-sized halo of
  $M_{200}=5\times10^{13}h^{-1}M_{\odot}$. More than $5\times10^6$
  realizations have been calculated for each case.  For indication,
  measured and model predicted flux ratios ($|\Rcusp|$ and $|\Rfold|$)
  of eight observed lenses are plotted as blue squares and cyan
  diamonds, respectively; measurement errors are also given. }
\label{fig:CuspViolationTotal}
\end{figure*}

We calculate the flux-ratio probability distributions with a total of
$5\times10^6$ realizations of generalised smooth lens potentials plus
(rescaled) subhalo populations from the Aquarius and the Phoenix
simulation suites.

The resulting flux-ratio probability distributions are presented in
Fig. \ref{fig:CuspViolationTotal}, where probability contours of
$P(>|\Rcusp|)$ for given $\Delta\phi$, $P(>|\Rcusp|)$ for given
$\theta/\Rein$ and $P(>|\Rfold|)$ for given $\theta_1/\Rein$ are
plotted. A small (large) probability $P$ means that it is less (more)
likely for a flux ratio, either $|\Rcusp|$ or $|\Rfold|$, to be larger
than a given value, at a given image configuration, described by
$\Delta\phi$, $\theta/\Rein$ or $\theta_1/\Rein$. The top panels show
the result from adopting smooth models; the middle panels show results
from using the smooth models plus a subhalo population hosted by a
Milky Way-sized halo of $M_{200}=10^{12}h^{-1}M_{\odot}$; the bottom
panels present results from taking a subhalo population hosted by a
group-sized halo of $5\times10^{13}h^{-1}M_{\odot}$. Note that these
distributions do not vary with the way that data are binned when using
a reasonable range of bin sizes.

To indicate the range of the observed flux ratios, on top of the
probability contours in the top panel of
Fig. \ref{fig:CuspViolationTotal}, measured $|\Rcusp|$ and $|\Rfold|$
of the eight lenses in our sample are plotted as blue squares,
together with measurement errors. Flux ratios that are predicted by
the lens model that best fits the image astrometry are also given,
plotted as cyan diamonds in the top panel.

It is important to realize that the forecasts shown in
Fig. \ref{fig:CuspViolationTotal} do not take into account the
magnification bias\footnote{Without correction for magnification bias,
  highly-magnified systems cannot be fairly sampled. Indeed these
  events only occupy a small fraction of the central caustic region in
  the source plane and thus would have lower weight in the statistical
  sample. However, due to the huge magnification effect they would be
  among the brightest detections in the Universe.}. Therefore the
predictions cannot be directly compared to the measurements made for
specific individual lenses; the calculations here are only aiming at
finding an allowed range and distribution of $\Rcusp$ and $\Rfold$.

%As can be seen, the generalised (smooth) host lens models can
%reproduce the majority of the model-predicted flux ratios with a
%probability of $P\gtrsim20-50\%$ (with the only exception of B1608);
%however, they can only reproduce the measured flux ratios with
%significantly lower probabilities.

The inclusion of substructures significantly broadens the flux ratio
distributions and increases the probabilities at larger values (for
close image configurations). As can be seen from the middle and bottom
panels of Fig. \ref{fig:CuspViolationTotal}, the {\sl values} of
$\Rcusp$ and $\Rfold$ measured for the observed lenses could indeed be
reproduced by adding CDM substructures to the generalised host lens
potential. The more massive the host halos are, the higher the
probabilities for having large $|\Rcusp|$ and $|\Rfold|$. This is
expected, as the number of subhalos increases with host halo mass (see
Sect. 3.2).

Adding substructures significantly changes the flux-ratio probability
distributions for the image triplets/pairs that have small
separations, but does not strongly affect the distributions on larger
scales. Such a variation behaviour confirms what one would expect from
local density perturbations: the image magnification and local
convergence satisfy $\mu\approx(1-2\kappa)^{-1}$ and thus
$\delta\mu/\mu\propto\mu \delta\kappa$ in the case of perturbation. A
close image configuration means that the image pairs must be located
close to the critical curves, where $\mu\rightarrow\infty$. Therefore
a tiny density fluctuation $\delta\kappa$ around the image positions
can cause a huge magnification fluctuation $\delta\mu$.

When a perturber is located near an image that is further away from
the critical curves (i.e., in the case of larger pair separations), it
is less efficient in altering the image magnification via a density
fluctuation. However, it could, if massive enough, shift the image to
a new position, where the magnification is different. In this case,
standard lens models (neglecting relatively massive perturbers if they
are not luminous enough to be seen) would have difficulties in fitting
the image positions. This is also referred to as ``astrometric
anomaly'' (e.g., \citealt{Chen2007astrometry}).

Due to the nature of the subhalo mass function, magnification
variations due to image position shifting (caused by relatively
massive subhalos) are expected to be less frequent than magnification
perturbation resulting from local density fluctuations (of lower-mass
subhalos), which will mainly occur for image pairs with small
separations around the critical curves. This is consistent with the
fact that only a small fraction of flux anomaly systems are also
reported to have astrometric anomalies
(\citealt{Biggs2004B0128,McKean2007B2045,Sluse2012COSMOGRAIL25}).

%In general, the effects of substructures on magnification
%perturbations and thus the flux-ratio probability distributions are
%more prominent for image triplets/pairs with smaller separations than
%for their large-separation counterparts. This makes systems that have
%smaller close-pair separations sensitive probes of CDM substructures
%via (radio) flux-ratio anomaly observations.

\section{Flux ratio probabilities of observed quadruple systems}
%: inclusion of substructures below the simulation resolution limit}

%%Using such generalised lens potentials is not aiming at reproducing
%%the specific image astrometry of individual lenses but only toward
%%finding an allowed range and distribution of $\Rcusp$ and $\Rfold$. 
%In this section, we further investigate the consequence of adding
%substructures to specific lens models of our observed sample and see
%how substructures perturb magnifications at the observed image
%positions for individual cases. The motivation for such an experiment
%is multifold:

%First, the application (as presented in Sect. 3) of generic smooth
%lens potentials has its limitations in at least the following two
%aspects: (1) without allowing for secondary lenses in the field, it
%cannot model systems with complex lens environments, e.g., satellite
%galaxies or nearby galaxy groups/clusters; (2) without considering
%magnification bias, it cannot fairly sample highly-magnified systems.

%Second, we also specifically aim at studying the perturbation effect
%from the vast amount of very low-mass subhalos. In order to
%disentangle different factors so that we understand the exact behavior
%of these very low-mass perturbers, we choose to study individual
%observed lenses to see how their flux ratio distributions are affected
%by the very low-mass subhalos.

In Sect. 3, we demonstrated that the {\sl values} of $\Rcusp$ and
$\Rfold$ measured at radio wavelengths for the quadruply lensed
systems could be reproduced by adding CDM substructures to generalised
lensing galaxy potential/mass distribution. Unfortunately, this approach
has its limitations in at least the following three aspects:

(1) Many lensing galaxies lie in rich environments
(\citealt{Momcheva2006LensEnv, Wong2011ShearEnv}). It is in general
necessary to account for the nearest lensing galaxy or group
explicitly in the model to reproduce the observed astrometry of those
systems. Accounting for these companions can modify the flux ratios at
the 30\% level in some observed systems, but this can only be
addressed on a case-by-case basis.

(2) A quantitative comparison between the forecast of Fig. 4 and the
observational data requires a proper understanding of the selection
effects of the sample, which should also be applied to the mocked data
from simulations. Unfortunately, this is not the case here.

(3) In any survey, the magnification bias plays an important role in
the selection of lens candidates, which enhances the probability to
observe highly magnified systems. Quantifying this bias is however not
an easy task. In addition, the flux ratios of those highly magnified
systems are more susceptible to the vast amount of very low-mass
subhalos, which were not included in the statistical calculations
presented in Sect. 3.

For all these reasons, we use in this section an alternative
methodology. We study the effects of CDM substructures in each
individual lens in our sample (instead of a generic population of
lenses) by adding substructures to a macro lens model that reproduces
the observed astrometry, and investigating the resulting flux ratio
distributions of images that closely resemble the observed
configurations.

We describe in Sect. 4.1 how we model the observed lenses, in
Sect. 4.2 how we add (to the macroscopic lens models) the subhalo
populations from the Aquarius and Phoenix simulation suites, and in
Sect. 4.3 how we model the subhalos that have masses up to three
orders of magnitude below the simulation resolution limit. The
ray-tracing method is described in detail in Sect. 4.4. In Sect. 4.5,
we carry out a case study using B2045+265 to investigate the
observational signatures of very low-mass subhalos and their
dependence on source sizes. Finally, the flux ratio probability
distribution for each of the observed systems are given in Sect. 4.6.

\subsection{Macro models of the observed lenses}

For the observed systems, we adopt a singular isothermal ellipsoid
plus a constant external shear $\gamma_{\rm ext}$ to fit only
astrometric measurements, i.e., positions of lensing galaxies, and
(VLBI/VLA) positions of lensed images. We do not use image flux ratios
to constrain the best lens models.  A second lens, being either a
satellite galaxy or a galaxy group, is also included in the model if
its optical/X-ray counterpart is seen in the same field (the induced
shear then may not be treated as constant). This second lens is
treated as a singular isothermal sphere (SIS).

Table \ref{tab:ObsSample-LensModel} lists parameters of our standard
lens models (SIE+$\gamma$+SIS) for systems in our sample. The
predicted $\Rcusp$ and $\Rfold$ of the close triple images (consistent
with those from the literature) are given in the parentheses in Col. 5
and Col. 7 of Table 1.

We note that two systems, i.e., B1608+656 and B1933+503, have been
excluded from such lens modelling and the discussions below, due to the
following reasons:

(1) Both lenses are spiral/discy galaxies
(\citealt{Fassnacht1996B1608, Sykes1998B1933}). This component has
however little effect on the image positions and is mostly constrained
by the flux ratios (\citealt{Maller2000HaloDisk, Moller2003Disc}). We
therefore do not expect that the simplified models adopted here are
appropriate for these two lenses.

(2) The images in the close triplets (in both lenses) are located far
away from the critical curve ($|\mu|<5$). As explained in Sect. 3,
local density perturbations are not expected to cause significant
magnification variation. Therefore CDM substructures are unlikely to
be responsible for the flux anomalies in these two cases.

\begin{table*}
   \centering
\caption{Best SIE+$\gamma$ (N$_{\rm lens} = 1$) and SIE+SIS+$\gamma$
(N$_{\rm lens} = 2$) models for our sample:}
\label{tab:ObsSample-LensModel}
\begin{minipage} {\textwidth}
\begin{tabular}[b]{l|c|c|c|c|c|c|c|c|c|c}\hline\hline  ~~~~~~~Lens &
~$z_{\rm l}$~ & $z_{\rm s}$ & $N_{\rm lens}$ &
~$\Rein$($\arcsec$)~ & ~~$e, \theta_{e} (\rm deg)$~~ & ~$\gamma,
\theta_{\gamma}(\rm deg)$~ &~$\Delta G($\arcsec$)$~ & $\chi^{2}$
(d.o.f.) & $ \chi^2_{\rm ima}, \chi^2_{\rm lens}$
   \\\hline
B0128+437$^1$ & 0.6 & 3.12 & 1  &  0.235 & 0.46, $-$27.72 & 0.213,41.17 &
0.006 & 0.4 (1)& 0.0, 0.4\\
%\hline
MG0414+0534$^1$ & 0.96 & 2.64 & 2  &  1.100, 0.181 & 0.22, 82.65 & 0.099,
$-$55.03 & 0.000 & 0.0 (0) & 0.0, 0.0\\
%\hline
B0712+472$^{\dagger2}$& 0.41 & 1.34  & 1  & 0.699 & 0.36, $-$61.8 &
0.076, $-$13.35  & 0.028 & 2.0 (1) & 1.95, 0.06 \\
%\hline
B1422+231$^{\clubsuit}$ & 0.34 & 3.62 & 2  &  0.785, 4.450 & 0.21, $-$57.62
& 0.091, 77.47 & 0.000 & 0.0 (1) & 0.0, 0.0 \\
%\hline
B1555+375$^{\spadesuit 3,4}$& 0.6 & 1.59 & 1 & 0.238 & 0.32,
81.26 & 0.143, $-$81.97 & 0.012 & 0.16 (1) & 0, 0.16\\
%\hline
%B1608+656$^1$ & 0.63  & 1.39 & 2  &  1.049, 0.094 & 0.84, 71.69  & 0.223,
%$-$10.70 & 0.000 & 0.0 (0) & 0.0, 0.0\\
%\hline
%B1933+503$^{\dagger \dagger \dagger \dagger 5}$& 0.76 & 2.63 & 1 & 0.517 &
%0.48, 43.51 & 0.032, 58.61 & 0.009  & 4.7 (1) & 1.2, 3.5\\
%\hline
B2045+265$^1$ & 0.87 & 1.28 & 2  &  1.101, 0.032 & 0.11, 29.09  & 0.203,
$-$67.07 & 0.000 & 0.0 (0) & 0.0, 0.0\\
\hline
\end{tabular}
\\ Notes: 
%($\dagger\dagger\dagger\dagger$) The model is based on positions of
%the lensing galaxy and the lensed images 1, 3, 4, 6 only. 
Col. 4 gives the total number of lenses included for modelling; Col. 6
provides the best-fitting amplitude and orientation of the
ellipticity; Col. 7 gives the external shear amplitude and the
position angle of the shear mass; Col. 8 provides the observed lensing
galaxy position with respect to the best-fitting lens position; Col. 9
gives the total $\chi^2$ of the best-fitting lens model; Col. 10
provides the independent contribution from the image and lens
astrometry to the total $\chi^2$. Note that flux ratios are not used
to constrain the models. ($\dagger$) Unrealistic lens models are
obtained when the nearby group positions of
\citet{Fassnacht02B0712Group} or \citet{Fassnacht2008Xray} are used;
therefore the group is not included in our lens
modelling. ($\clubsuit$) This model uses the X-ray centroid of the
group by \citet{Grant2004Xray}. ($\spadesuit$) We assume $(z_{\rm l},
z_{\rm s}) = (0.6, 1.59)$ and use the galaxy position from CASTLES,
$(\Delta_{\rm gal} {\rm RA},~\Delta_{\rm gal} {\rm DEC})$ = ($-$0.185,
$-$0.150)$\pm$0.03$\arcsec$ with respect to image A. References: (1)
\citealt{Sluse2012COSMOGRAIL25}; (2) \citealt{Jackson2000}; (3)
\citealt{Marlow1999B1555}; (4) CfA-Arizona Space Telescope Lens Survey
(CASTLES, see http://cfa-www.harvard.edu/castles). %(5)
\citealt{Cohn2001B1933}.
\end{minipage}
\end{table*}

\subsection{Adding Aquarius and Phoenix subhalos to the macroscopic lens models}
In order to maintain the macroscopic critical curve, we renormalize
the macroscopic convergence $\kappa_{\rm mac}$ by a factor of
(1-$\kappa_{\rm sub}$), where $\kappa_{\rm sub}$ is the convergence
from the total amount of subhalos (including the very low-mass ones)
projected in the central region. We then add to the best-fitting
macroscopic lens potentials the simulated subhalo populations at above
$10^7h^{-1}M_{\odot}$ taken from the snapshots with redshifts closest
to the observed lens redshifts.

In this section, for each of the observed lenses, we rescale the
simulated subhalo populations to match a host halo whose inner
velocity dispersion (estimated by $1/\sqrt{2}$ of the peak velocity as
first-order approximation) is equal to the one constrained by the
best-fitting SIE model. The masses of the rescaled host halos that are
supposed to host the observed lenses range from
$10^{12}h^{-1}M_{\odot}$ (for B0128+437 and B1555+375) to
$2.5\times10^{13}h^{-1}M_{\odot}$ (for B2045+265).

For each lens, we draw $\sim$250 projections from each of the
simulated host halos, i.e., $\sim3800$ projections in total from all
fifteen halos in the two simulation suites. For each realization
(including adding very low-mass subhalos in Sect. 4.3), multiple
candidate source positions within a radius of $0.01\arcsec$ around the
model-constrained source position (with respect to the caustic) were
searched for close triple images. Here we further adopt a selection
criteria so that only systems that best resemble the observed image
geometry would be chosen. The criteria are applied to the
configuration parameters $\Delta\phi$, $\theta/\Rein$ and
$\theta_1/\theta$ of the close triplets in each simulated system. We
require the relative differences between the simulated and the
observed quantities to be no larger than 10\%:
\begin{equation}
\begin{array}{c} 
\displaystyle \left|\frac{(\Delta\phi)_{\rm
    sim}}{(\Delta\phi)_{\rm obs}}-1\right|\leqslant0.1,\\ 
\displaystyle \left|\frac{(\theta/\Rein)_{\rm sim}}{(\theta/\Rein)_{\rm
    obs}}-1\right|\leqslant0.1, \\ 
\displaystyle \left|\frac{(\theta_1/\theta)_{\rm
    sim}}{(\theta_1/\theta)_{\rm obs}}-1\right|\leqslant0.1.
\end{array}
\label{eq:selection}
\end{equation}

\subsection{Model subhalos beyond the CDM simulation resolution limit}
As an important complement to current studies, we investigate the
lensing effects from subhalos with masses between $10^4 h^{-1}
M_{\odot}$ and $10^7 h^{-1} M_{\odot}$. For simplicity and clarity,
three specific masses fixed at $m_{\rm sub} = 3\times 10^4 h^{-1}
M_{\odot}$, $3\times 10^5 h^{-1} M_{\odot}$ and $3\times 10^6 h^{-1}
M_{\odot}$ are used for the three different mass decades in
question. Assuming that the subhalo mass function and their profile
parameters follow power-law functions of mass, we extrapolate the
spatial distribution $\eta(R)$ and density profiles $\rho(r)$ of
subhalos from the Aquarius and the Phoenix simulations to these very
low-mass scales considered here. Three different source radii $r_{\rm
  s}$ of 1pc, 3pc and 5pc, reflecting the different sizes of the
emission regions of lensed quasars, are also applied to investigate
the source size dependence.
%The default spatial distribution $\eta(R)$ and density profiles
%$\rho(r)$ of these very low-mass subhalos are reasonable
%extrapolations based on the higher-mass subhalos from the Aquarius and
%the Phoenix simulations.

\subsubsection{Projected number density distribution}
The halo-centric distribution of the projected number densities
$\eta_{m}(R)$ of the low-mass subhalos, where $m=4,~5,~6$ for the
three different mass bins studied here, are extrapolated from that of
their higher-mass counterparts -- the resolved subhalos from the
rescaled Aquarius and Phoenix simulations.  From
Fig. \ref{fig:SubProjDistribution}, it can be seen that the subhalo
number density $\eta(R)$ at a given mass decade almost remains
constant in the inner region of their host. This density increases by
a factor of 10 each time when subhalo masses decrease by one
decade. We therefore model the projected number densities $\eta_{m}$
of low-mass subhalos by: $\eta_{m}=\eta_{78} \times 10^{(7-m)}$, where
at $\eta_{78}$ is the projected number density of subhalos of
$10^{7\sim8} h^{-1} M_{\odot}$. The projected positions of the
low-mass subhalos are then randomly distributed in the lens plane,
according to their projected number densities $\eta_{m}$.

\subsubsection{Density profiles}
\begin{figure}
\centering \includegraphics[width=7cm]{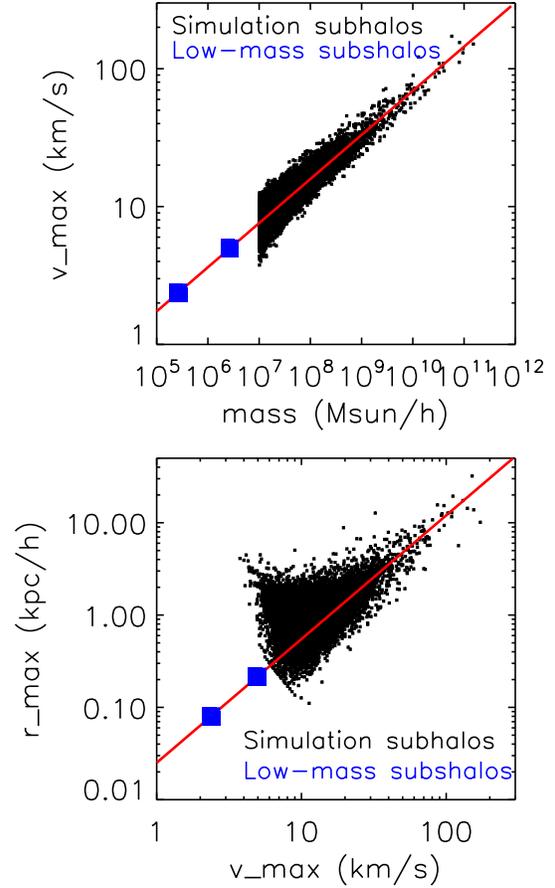}
\caption{The extrapolation of the $V_{\rm max} - m_{\rm sub}$ and
  $r_{\rm max} - V_{\rm max}$ relations using subhalos (black dots)
  from one of the level-two Aquarius halos. The red lines are given by
  Eq. (\ref{eq:MassVmaxRmax}). The blue symbols represent the adopted
  values according to the relation for subhalos at $3\times10^{5}
  h^{-1} M_{\odot}$ and $3\times10^{6} h^{-1}
  M_{\odot}$.} \label{fig:VmaxRmaxMsubRelation}
\end{figure}

As in Sect. 3, we assume subhalos to be modelled by Einasto profiles
(Einasto 1965) with slope parameter $\alpha=0.18$. The other two
parameters that are required to fix the profile are $V_{\rm max}$ and
$r_{\rm max}$, both of which are measured for the resolved subhalos in
the Aquarius and Phoenix simulations (see Sect. 3.2.3). The sets of
parameters $V_{\rm max}$ and $r_{\rm max}$ for the low-mass subhalos
studied here are obtained by extrapolating the $V_{\rm max} - m_{\rm
  sub}$ and $r_{\rm max} - V_{\rm max}$ relations that exist, albeit
not tight, for their higher-mass
counterparts. Fig. \ref{fig:VmaxRmaxMsubRelation} shows an example of
the extrapolation using subhalos from one of the level-two Aquarius
halos. As the fitting formula for the $V_{\rm max} - m_{\rm sub}$ and
$r_{\rm max} - V_{\rm max}$ relations change little at redshift $z<1$,
we take the following uniform fitting expressions:
\begin{equation}
\begin{array} {l}
\displaystyle v_{\rm max} = 3.6~\kms \left(\frac{m_{\rm sub}}{10^6~h^{-1}M_{\odot}}\right)^{0.32} \\
\displaystyle r_{\rm max} = 0.55~\rm kpc/h \left(\frac{v_{\rm max}}{10~\kms}\right)^{1.34}
%\log v_{\rm max} [{\rm km/s}] = 0.32 \log m_{\rm sub} [h^{-1}M_{\odot}] - 1.36 \\
%\log r_{\rm max} [{\rm kpc/h}] = 1.34 \log v_{\rm max} [{\rm km/s}] -1.6
\end{array}
\label{eq:MassVmaxRmax}
\end{equation}

\begin{figure*}
\centering \includegraphics[width=5.5cm]{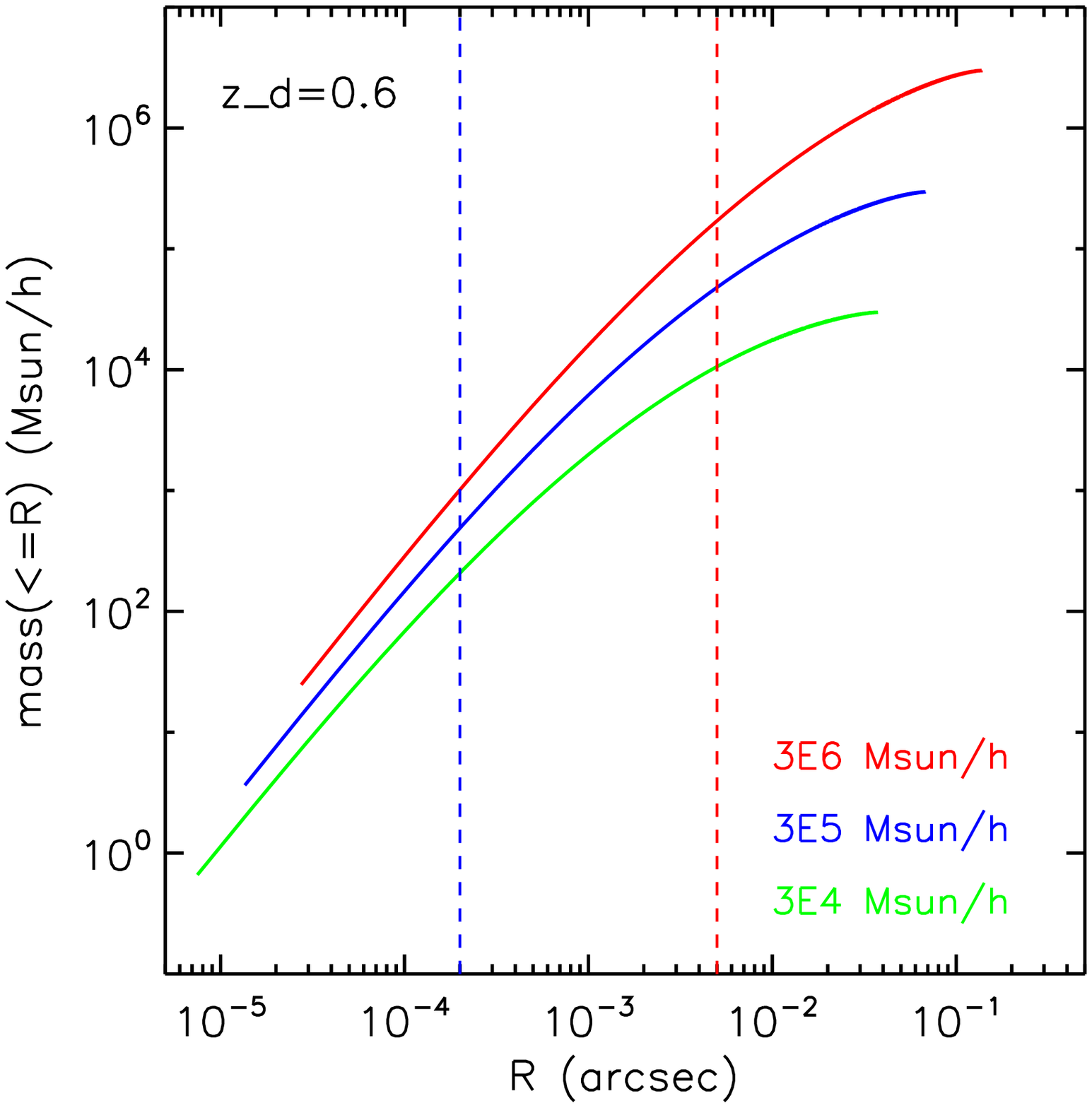}
\centering \includegraphics[width=5.5cm]{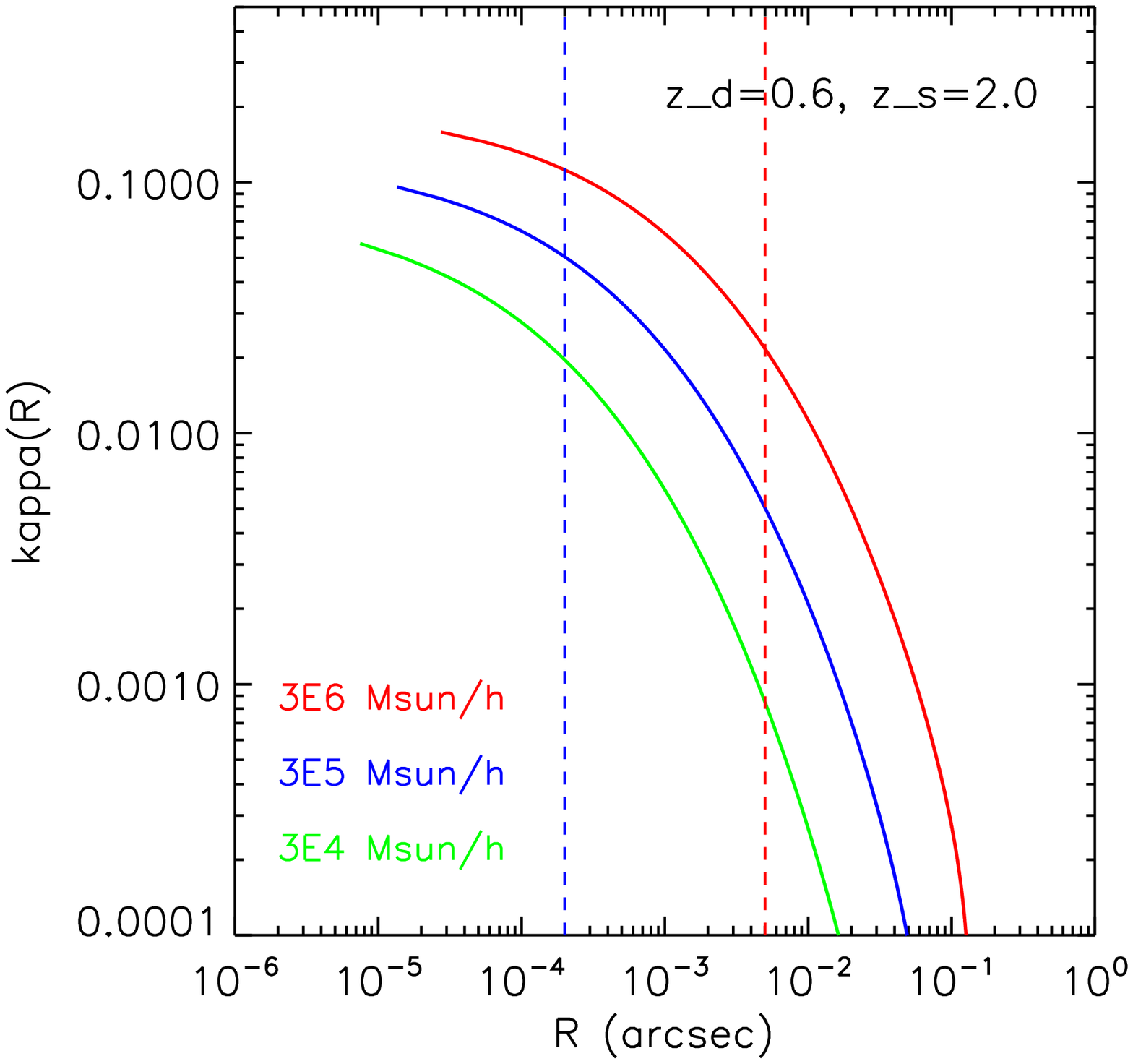}
\centering \includegraphics[width=5.5cm]{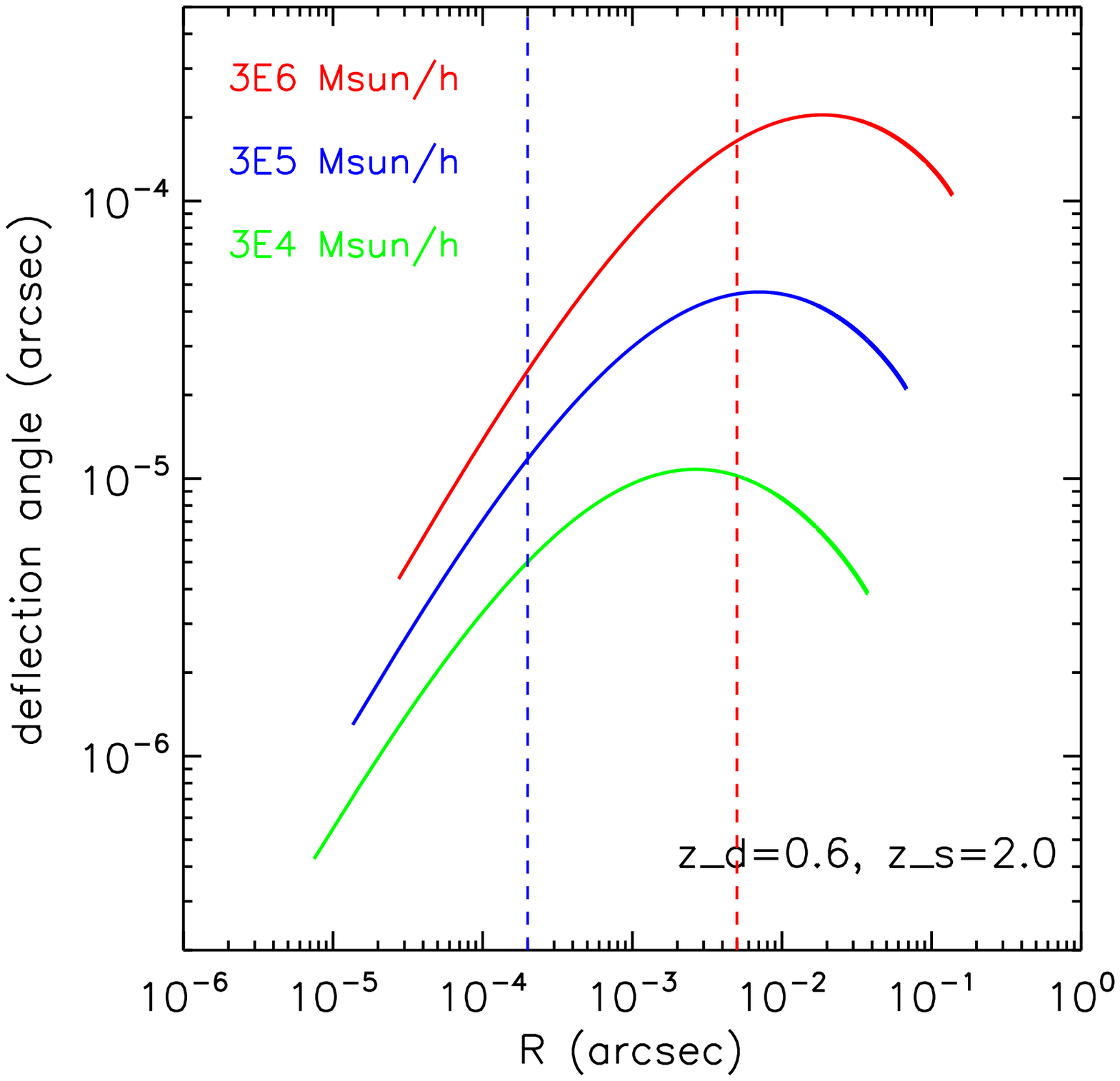}
\caption{The enclosed mass (left), convergence (middle) and deflection
  angle (right) distributions as a function of radius for
  Einasto-profiled subhalos at $3\times10^{4} h^{-1} M_{\odot}$
  (green), $3\times10^{5} h^{-1} M_{\odot}$ (blue) and $3\times10^{6}
  h^{-1} M_{\odot}$ (red). The dashed red (blue) vertical line
  indicates the lens-plane mesh resolution used in the calculations
  presented in Sect. 3 (Sect. 4), which ensures that the least massive
  subhalos in question can be resolved by a few to ten pixels at radii
  where their half masses and peak deflection angles are
  reached. } \label{fig:EinastoProfileLowMassRes}
\end{figure*}

Each subhalo will be truncated at a radius $r_{\rm t}$, within which
the enclosed mass is equal to the given subhalo mass, i.e.,
$m(\leqslant r_{\rm t})=m_{\rm sub}$. At an assumed lens redshift
$z_{\rm l}=0.6$, subhalos of $m_{\rm sub} = 3\times 10^4
h^{-1}M_{\odot}, ~3\times 10^5 h^{-1} M_{\odot}$ and $3\times 10^6
h^{-1}M_{\odot}$ are truncated at $r_{\rm t}=0.03\arcsec,~0.06\arcsec$
and $~0.12\arcsec$,
respectively. Fig. \ref{fig:EinastoProfileLowMassRes} shows the
enclosed mass profile, convergence profile and the distribution of
deflection angle as a function of projected radius.

With the Einasto-profile parameters fixed, lensing properties can be
calculated at any given position in the lens plane. Once again we vary
$r_{\rm max}$ by a factor of 0.71 and 0.58 from the default value (so
that the overdensity $\delta\equiv(v_{\rm max}/r_{\rm max})^2$ varies
by a factor of 2 and 3) and repeated the same calculation. We verify
that such variations do not bring marked difference in the flux-ratio
distributions.

\subsection{Ray-tracing for the magnification calculation}

The lensing effect of the low-mass subhalos strongly depends on the
size of the emission region of the source, i.e., the smaller the
latter is, the stronger the former would be. Our numerical approach
needs to reproduce image magnifications for various source sizes.

To ensure that the regions of interests will be sampled with enough
resolution, we use a finer lens-plane mesh with a resolution of
0.0002$\arcsec$/pixel that covers the observed image triplets to
calculate the lensing properties (i.e., the first- and second-order
derivatives of the lens potentials) of the main lens and of the
subhalos. Again image positions and magnifications of a point source
are found through a Newton-Raphson iteration method. To find the image
magnifications of a finite-sized source at $\vec{\beta}^{\star}$ with
radius $r_{\rm s}$, we start casting rays from the grid points
$\vec{\theta}$ of the regular lens-plane mesh to the source plane
according to the lens equation, all the resulting source positions
$\vec{\beta}$ that satisfy $|\vec{\beta}-\vec{\beta}^{\star}|\leqslant
r_{\rm s}$ are picked out. Their lens-plane counterparts map out three
groups that correspond to the triple images of the given finite-sized
source. The image magnification $\mu^{\star}$ of each image is then
given by:
\begin{equation}
\mu^{\star} =
\frac{\Sigma_{i}\delta\theta^2}{\Sigma_{i}\delta\beta_i^2}=
%\frac{\Sigma_{i}\delta\theta^2}{\Sigma_{i}\delta\theta^2/\mu_{i}}=
\frac{\Sigma_{i}\delta\theta^2}{\pi
  r_{\rm s}^2},
\end{equation}
where $\delta\theta^2$ is the uniformly-sampled finite area element in
the image plane, and $\delta\beta^2=\delta\theta^2/\mu_{i}$ is the
corresponding area element in the source plane. The summation
$\Sigma_i$ is over all the test positions $\vec{\theta}_{i}$ 
%(with magnifications given by $\mu_{i}$ and generated around a given
%image in the close triplet)
that are mapped to the source plane where
$|\vec{\beta}_{i}-\vec{\beta}^{\star}|\leqslant r_{\rm s}$.

\subsection{Impact of very low-mass subhalos and the finite source effect} 

In this subsection, we investigate whether subhalos below the
resolution limit, i.e., $\sim10^7 h^{-1}M_{\odot}$, can still produce
significant flux ratio anomalies. For this purpose we perform the case
study of B2045+265: we take the macroscopic lens model and image
geometry and calculate the perturbation effects from inclusion of
subhalos of $m_{\rm sub}=3\times10^4 h^{-1}M_{\odot}$, $3\times10^5
h^{-1}M_{\odot}$ and $3\times10^6 h^{-1}M_{\odot}$. In particular, we
study several cases with different combinations of subhalo properties
and source sizes. In each case, we repeat the magnification
calculation 2000 times to obtain different realizations of subhalo
spatial distributions. The statistical distributions of image
magnification ratios are case-dependent, and thus unveil how the
perturbation effects from the very low-mass subhalos depend on their
masses and source sizes.

\begin{figure}
\centering
\includegraphics[width=8cm]{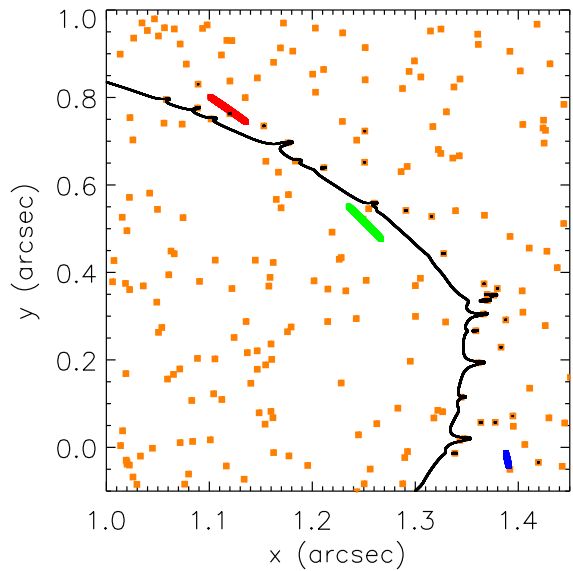}
\caption{Tangential critical curve of B2045+265, small-scale wiggles and
  isolated local critical curves are induced by subhalos of
  $10^{5\sim6} h^{-1}M_{\odot}$ (orange). Red, green and blue regions
  are the close triplets of the theoretical source with a finite
  radius of 5pc. }
\label{fig:CCB2045}
\end{figure}

\begin{figure}
\centering
\includegraphics[width=8cm]{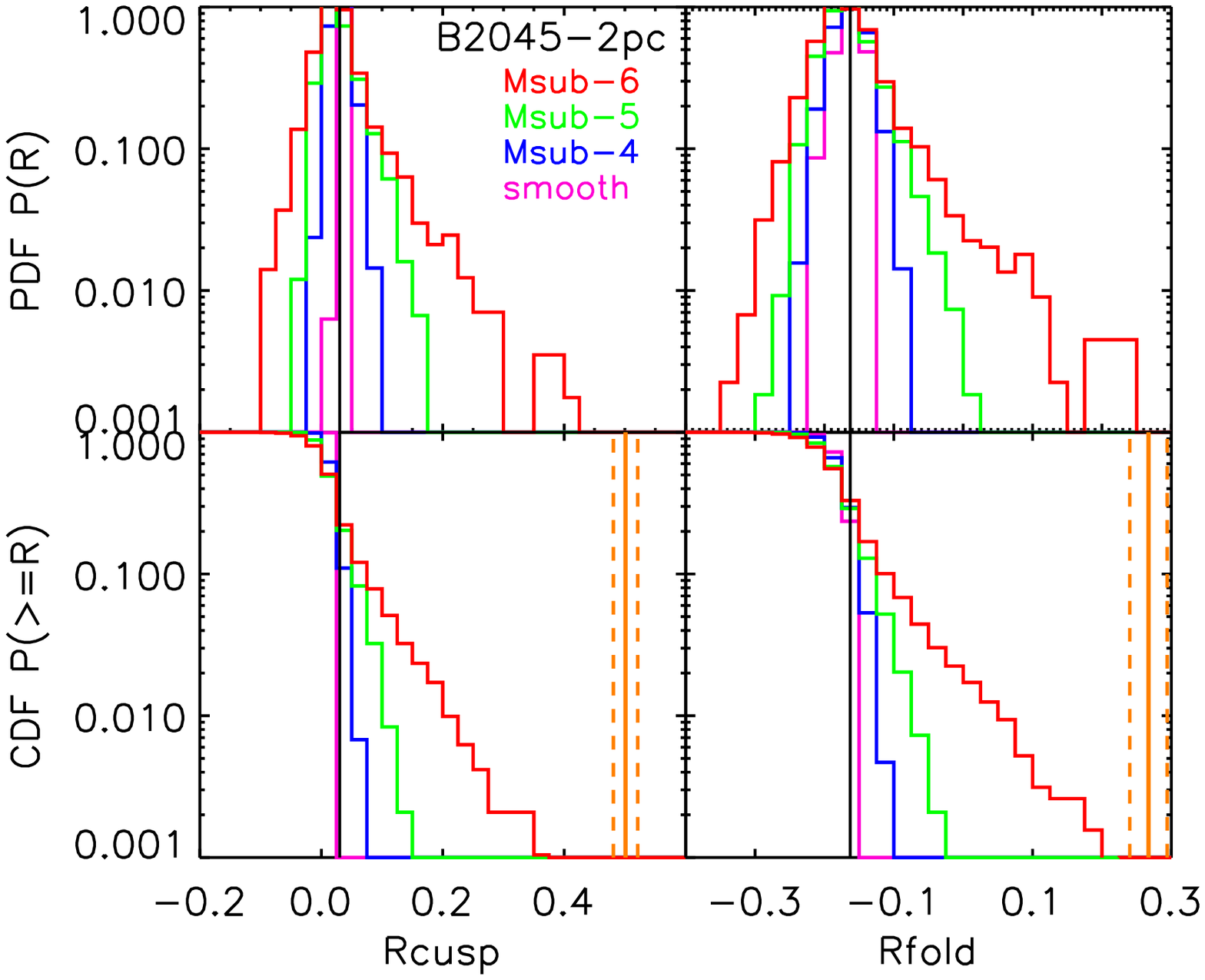}
\includegraphics[width=8cm]{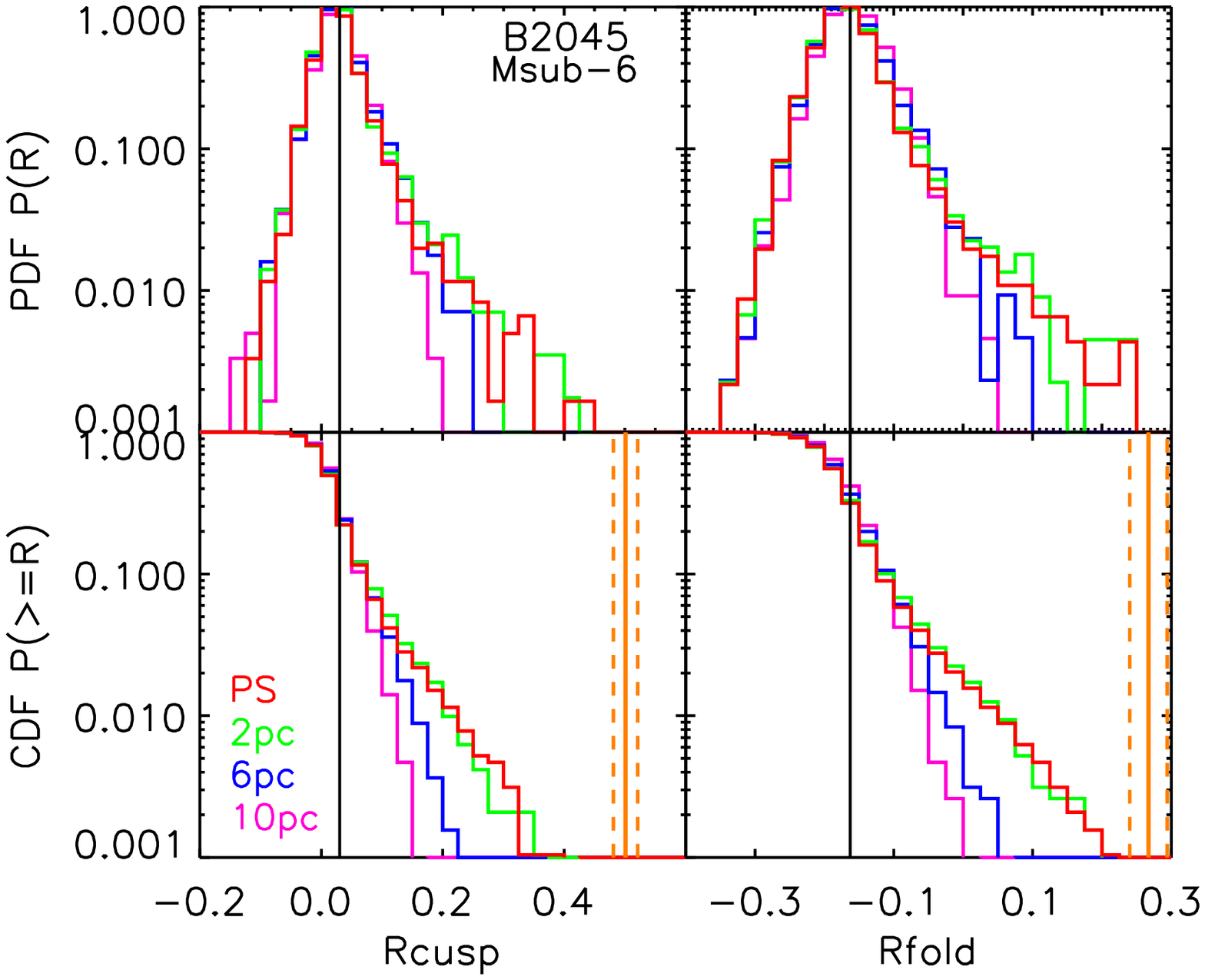}
\caption{The differential and cumulative probability distributions of
  $\Rcusp$ (left) and $\Rfold$ (right), calculated under different
  scenarios. The top panel presents results for a finite-sized source
  of 1pc in radius, the included subhalo masses are at
  $3\times10^4h^{-1}M_{\odot}$ (green), $3\times10^5h^{-1}M_{\odot}$
  (blue) and $3\times10^6h^{-1}M_{\odot}$ (red) as well as in the
  absence of substructures (pink). The bottom panel shows results for
  subhalos at $3\times10^6h^{-1}M_{\odot}$ but assuming a point source
  (red), a finite-sized source of 1pc (blue), 3pc (green) and 5pc
  (pink) in radii. Orange vertical lines indicate the measured flux
  ratios (and the uncertainties) for
  B2045+265. }\label{fig:RcuspRfoldTestLens7}
\end{figure}

As $\delta\mu/\mu\propto\mu \delta\kappa$, at around the main critical
curve, even a small mass fluctuation $\delta\kappa$ (from subhalos)
could modify the shape of the critical curve. This is demonstrated in
Fig. \ref{fig:CCB2045}, where an example using B2045+265 is given. In the
case of very low-mass subhalos, localized critical lines could form
around these perturbers on milli-arcsecond (mas) to sub-mas
scales. When an image that is located near the main critical curve
happens to cover these localized critical lines, the brightness of the
image can be significantly enhanced if the image size is on similar
scales ($\la0.001\arcsec$).

Fig. \ref{fig:RcuspRfoldTestLens7} shows the differential and
cumulative probability distributions of $\Rcusp$ and $\Rfold$,
calculated under different scenarios. The top panel presents results
for a finite-sized source of a fixed radius but using three different
subhalo masses. The bottom panel shows results for subhalos at a fixed
subhalo mass but assuming a point source and a finite-sized source of
different radii.

It is clearly seen that the perturbation effects on the flux ratios
become significant with increasing subhalo masses $m_{\rm sub}$, even
though the number densities $\eta_{\rm sub}$ decrease. Convergence
tests with different lens-plane resolution at $0.0001\arcsec$,
$0.0002\arcsec$ and $0.0005\arcsec$ per pixel confirmed that such
numerical results are genuine and not due to insufficient
resolution. This is expected, as explained in Xu et al. (2009),
because for simulated subhalos or point masses, the lensing
cross-section $\sigma$ of a subhalo can be approximated by
$\sigma\propto m_{\rm sub}^\alpha$, where $\alpha$ is a positive
index, thus rendering the total lensing cross-section dominated by
massive subhalos. For this reason, when calculating the flux-ratio
probability distributions for each of the observed lenses, we safely
neglect subhalos below $10^5h^{-1}M_{\odot}$ so that the computational
expense stays low.

On the other hand, when the subhalo mass is fixed, we see that the
smaller the source size is, the more extended the distribution tail
becomes; point sources yield the most significant extension at large
flux ratios. In the next subsection, we therefore only present results
for each observed system under the point source assumption to achieve
an upper bound of possible substructure perturbation.

Another very interesting feature seen from
Fig. \ref{fig:RcuspRfoldTestLens7} is that the distribution is skewed
towards larger $\Rcusp$ (and $\Rfold$).
%there are larger probabilities for the ratios to be larger than the
%average values.
Such an asymmetric distribution is rooted in the tendency that saddle
(minimal) images become fainter (brighter) where clumpy substructures
are present near the image positions. Such a behaviour from low-mass
subhalos is similar to that shown in the bottom panel of Fig. 3 in
\citet{SW2002apj}.

\subsection{Results for individual lenses}

\begin{figure*}
\centering
\includegraphics[width=5.5cm]{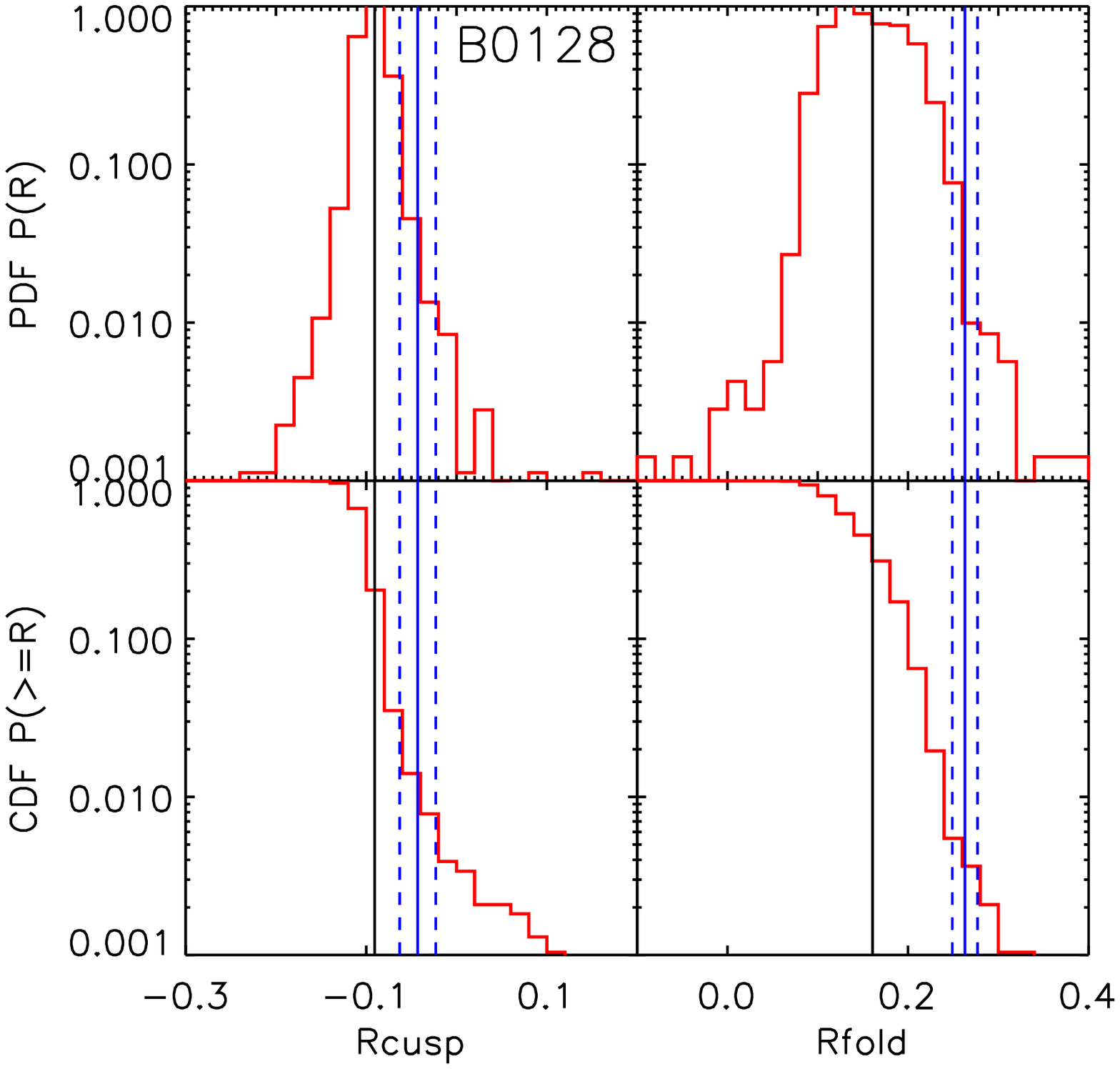}
\includegraphics[width=5.5cm]{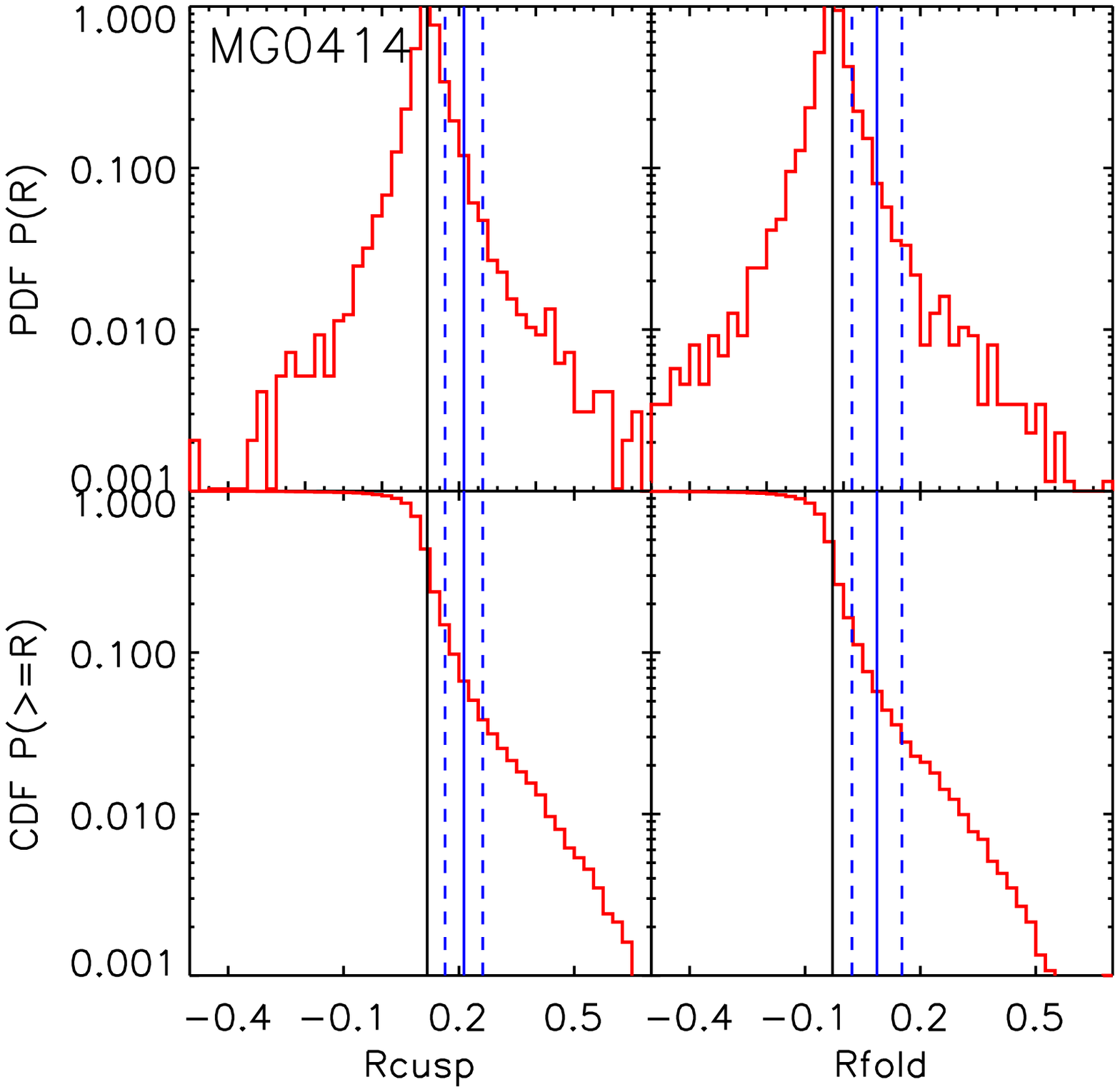}
\includegraphics[width=5.5cm]{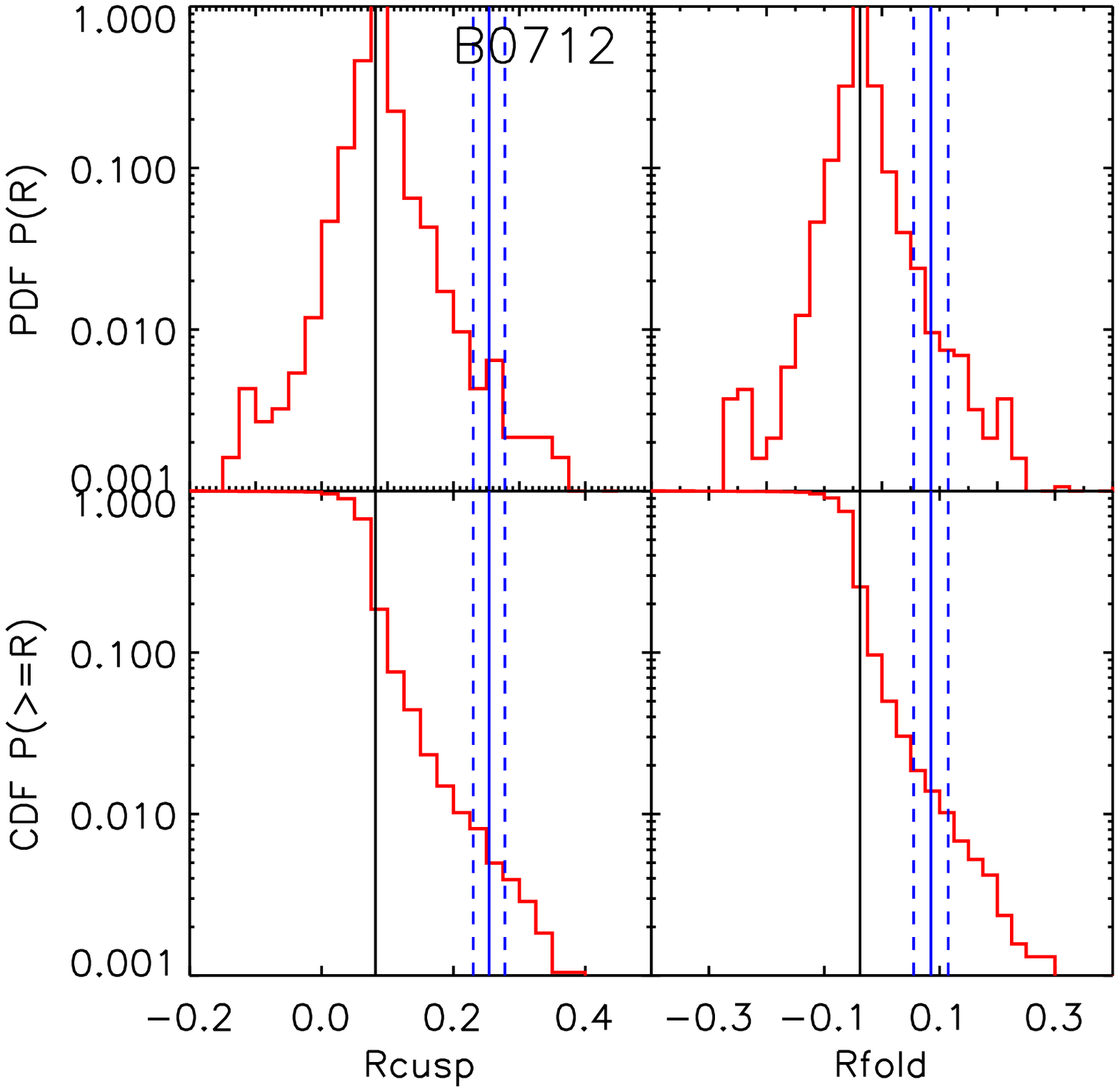} \\
\includegraphics[width=5.5cm]{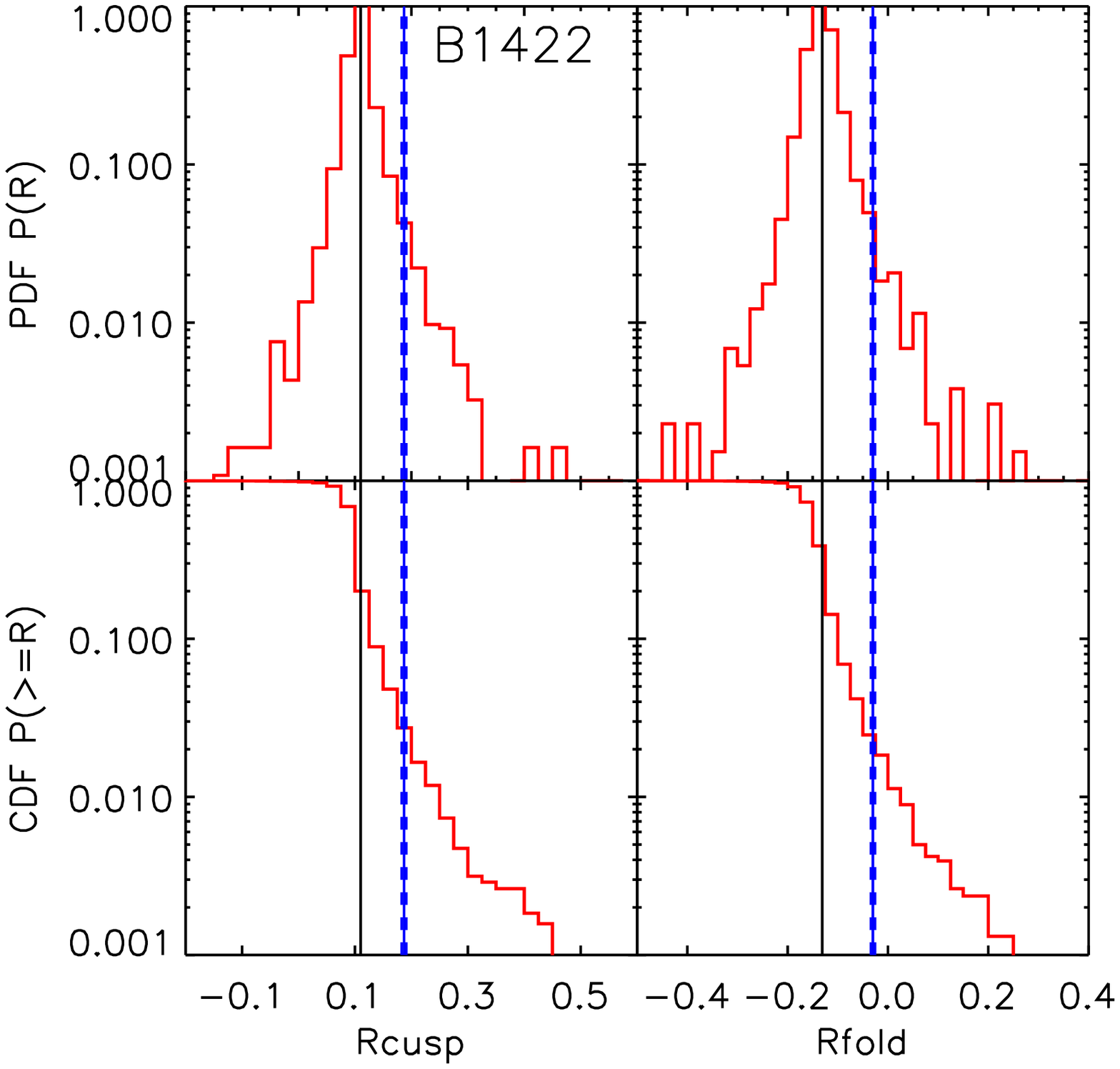}
\includegraphics[width=5.5cm]{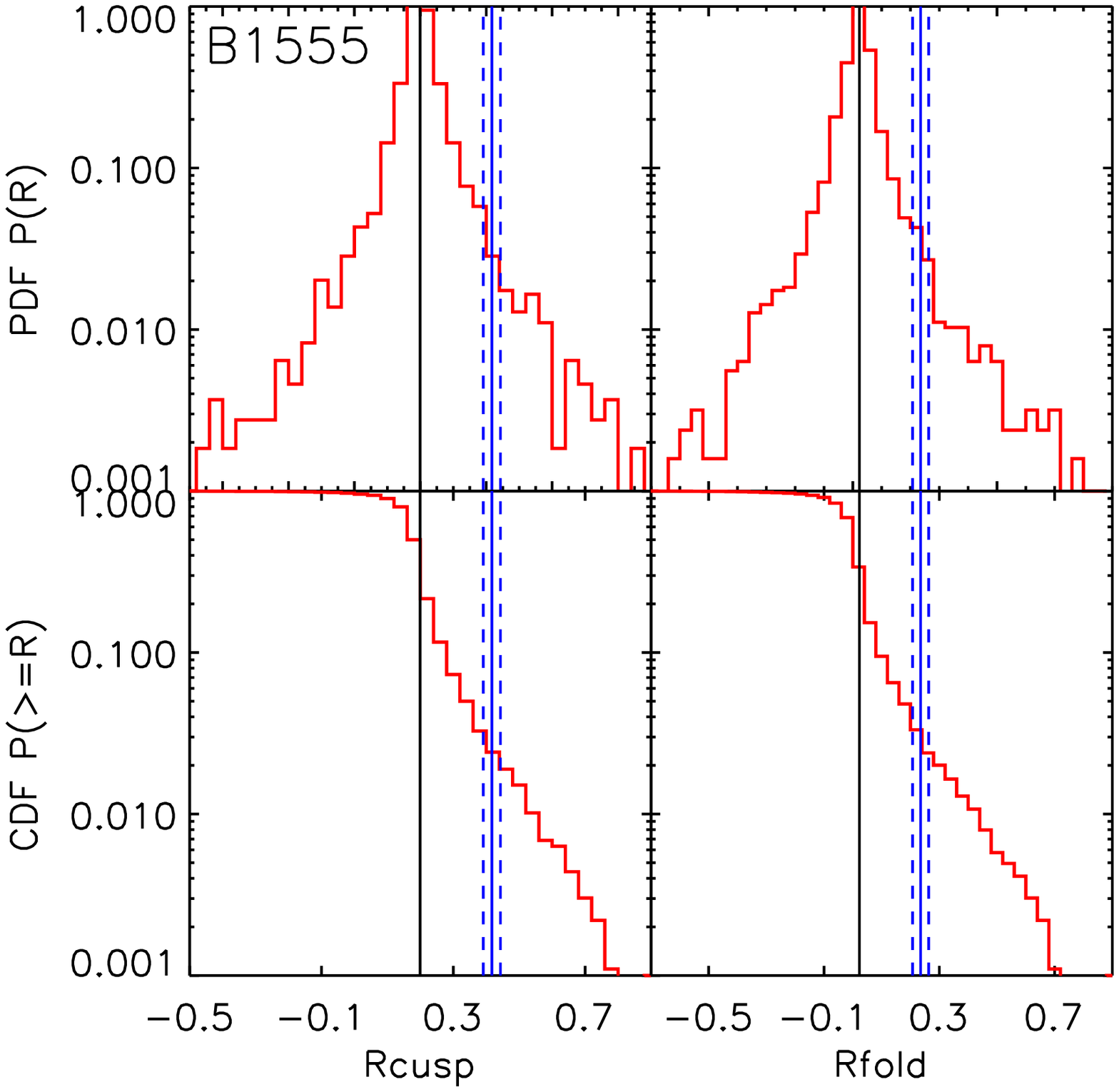}
\includegraphics[width=5.5cm]{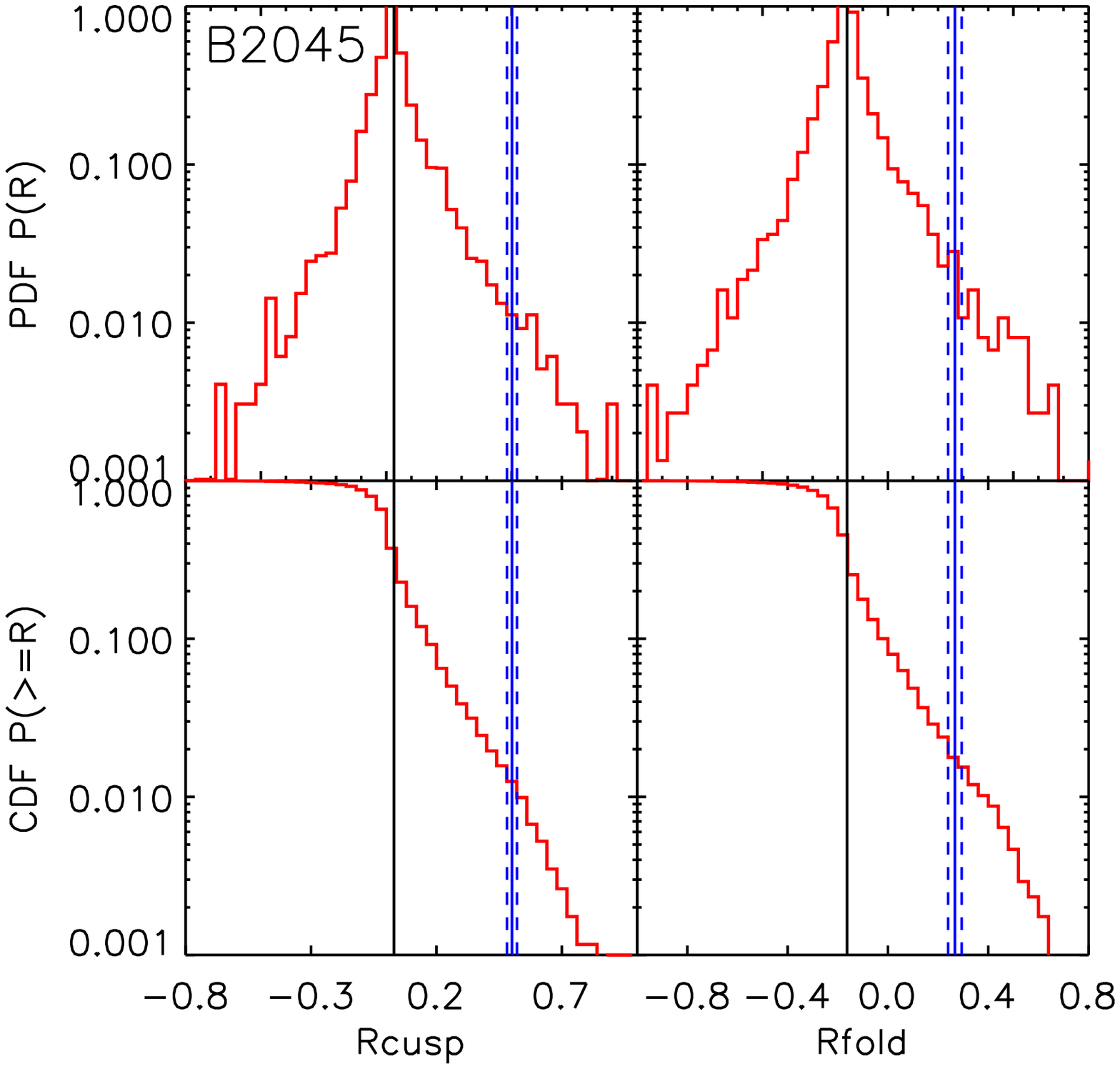}
\caption{Flux ratio probability distributions (same as
  Fig. \ref{fig:RcuspRfoldTestLens7}) for selected realizations that
  most resemble each observed system. Red solid lines represent
  results from including CDM substructures above
  $10^5h^{-1}M_{\odot}$. Measured and predicted flux ratios are
  indicated by the blue and black vertical lines, respectively. Dashed
  lines indicate errors on the measurements. }
\label{fig:ViolationIndivSystems1}
\end{figure*}

Below we present the flux-ratio probability distributions for each of
the observed lens systems, calculated using their observed specific
image configurations and their own lens models, plus CDM substructures
above $10^5h^{-1}M_{\odot}$. Fig. \ref{fig:ViolationIndivSystems1}
shows the probabilities to have $\Rcusp$ and $\Rfold$ larger than the
observed values in each systems. 

As can be seen, such probabilities are about $5\%-20\%$ (taking into
account the large measurement uncertainties) for MG0414+0534. For the
rest of the lenses in our sample, the probabilities are only
$1\%-4\%$. In principle, the close image geometries of these systems
should make their flux ratios more susceptible to density
perturbations (e.g., from CDM substructures). However, such
percent-level probabilities indicate that there must be other sources
for the mismatch between the measured and model predictions. For
example, VLBI observations already showed evidence of scatter
broadening in B0128+437 (Biggs et al. 2004). For B0712+472, a galaxy
group has been identified on its line of sight (Fassnacht \& Lubin
2002, Fassnacht et al. 2008). We were unsuccessful in accounting for
this group in the smooth lens model (Table
\ref{tab:ObsSample-LensModel}) due to its uncertain X-ray
centroid. The lens model for B1555+375 might also not be optimal, as
the position angles of the ellipticity and of the external shear are
nearly orthogonal; the HST images also suggest a flattened
morphology. All these strongly indicate a possibly missing ingredient
in the lens model.
%We note that our approach to estimate the host halo mass for an
%observed lens could be biased, while the flux-ratio probabilities
%increase with the host halo mass (and subhalo abundance), we
%... result doens't change!
In the next section, we discuss other possible reasons to account for
the discrepancy.

\section{DISCUSSION AND CONCLUSIONS}

\subsection{The contribution from CDM substructures}

In Sect. 3 we see (from the bottom panel of Fig. 4) that the inclusion
of CDM substructures reproduces those large values of $\Rcusp$ and
$\Rfold$ seen in observations. Indeed, among the observed lenses in
our sample, McKean et al. (2007) found that a lens model that
incorporates an observed dwarf satellite (the luminous counterpart of
a dark matter subhalo) could reproduces all image positions as well as
the flux ratios for B2045+265. \citet{MacLeod2012MG0414sub} also
showed that the observed flux ratios in MG0414+0534 can be reproduced
by adding a substructure of $\sim 10^7 M_{\sun}$ close to image
$A2$. Again with detailed lens modelling, Nierenberg et al. (2014)
found a better fit to the image astrometry as well as the flux ratios
when adopting a lens model that includes a perturbing mass of
$10^{7\sim8} M_{\sun}$ around image $A$. Other works, e.g.,
\citet{Bradac2002B1422}, Dobler \& Keeton (2006) and
\citet{Fadely2012HE0435Sub}, also found that the inclusion of a local
perturbation with mass of $10^{5\sim8} M_{\sun}$ can always help to
explain the image flux ratios measured at longer wavelengths.

On the one hand, the flux ratios can always be ``fixed'' by adding
local density perturbations to the smooth lens potential that
reproduce the observed macroscopic image positions. However, it is
interesting that coincidentally the required masses of the added
perturbers happen to be within a range that is predicted for abundant
low-mass CDM subhalos that survive the tidal destruction during galaxy
formation.

But on the other hand, when deploying a theoretical population of the
CDM subhalos from cosmological simulations, we find that 
%the chance is small for a CDM subhalo in a relevant mass range to
%coincide (in projection) with the images: the probabilities to
%reproduce these observed $\Rcusp$ and $\Rfold$ are only of a few
%percent.
%It is intriguing and actually puzzling that
even for systems (like B1555+375 and B2045+265) that are more susceptible to
local density perturbations,
%like B1555 and B2045, where the close triplets/pairs are located so
%closely around the critical curve that even a tiny density
%perturbation (at the image positions) would induce significant
%magnification fluctuations, 
the probabilities to reproduce values of $\Rcusp$ and $\Rfold$ larger
than the measured values are only at per cent level.
%, assuming that CDM substructures are the only source of density
%perturbation.
This strongly indicates that there are other culprits for the radio
flux-ratio anomalies. In the next subsection, we present these other
possibilities. % that could account for the discrepancy.

We mention in passing that CDM substructures could not only affect the
radio flux ratios of a multiply-imaged quasar, but also leave imprints
on the surface brightness distribution of a lensed galaxy. Through the
detection and modelling of these image distortions, one can also
constrain the level of density perturbations in a mass range of
$10^{6\sim9}h^{-1}M_{\odot}$ in a lensing galaxy
(\citealt{VK2009substatistics}). This has already been put into good
practice by e.g., \citet{Vegetti2012Nature, Vegetti2014SubFrac} on the
SLACS lenses using high resolution HST and Keck adaptive optics
imaging. The resulting CDM substructure fraction is in consistent with
the ones derived from $N$-body simulations (see Sect. 3.2.2). More
high-resolution images of lensed dusty star-forming galaxies will also
soon be available from ALMA, which can also be used to constrain CDM
substructures via the induced image distortion
(\citealt{Hezaveh2013CDMSubALMADustLens}).

\subsection{Other culprits for radio flux-ratio anomalies}

%It is intriguing and actually puzzling that even for systems like
%B1555 and B2045, where the close triplets/pairs are located so closely
%around the critical curve that even a tiny density perturbation (at
%the image positions) would induce significant magnification
%fluctuations, the probabilities to reproduce $\Rcusp$ and $\Rfold$
%larger than the measured values are only at per-cent level, assuming
%that CDM substructures are the only source of density
%perturbation. This strongly indicates that there could be other
%culprits for the radio flux-ratio anomalies.

\citet{Xu2012LOS} investigated the effects from CDM halos along the
line-of-sight to a lensed quasar. Comparing Fig. 9 therein with
Fig. \ref{fig:CuspViolationTotal} here, it can be seen that the
contribution of these interlopers can be as important as that of the
intrinsic CDM substructures within the lensing galaxy \citep[also,
  e.g.,][]{Metcalf2005a, Metcalf2005b, Miranda2007}. However even
factoring in the effects from line-of-sight perturbers, the gap
between the observed flux ratios and model predictions still remains.

In \citet{Dandan2010AqII}, three types of substructures other than
bound CDM subshalos were investigated, i.e., satellite galaxies,
globular clusters (GC) and satellite streams, which were found to
contribute little to solving the radio flux-ratio anomaly
problem. However the adoption of an empirical Milky-Way GC population
has a caveat. As also pointed out in their discussion, massive
elliptical galaxies are known to host more GCs than their spiral
counterparts (\citealt{FMM1982,Harris1991,Harris1993,West1993}). GCs
are typically of mass $10^{5\sim6}M_{\odot}$ and have the most compact
density profiles among all known types of galactic substructures. We
would hence like to point out the possibility of massive elliptical GC
populations to be an extra source of relevant perturbations for our
problem.

Apart from GCs, baryonic substructures may also exist at above the
$10^6M_{\odot}$ level. When a small halo merges with a bigger halo,
and later on becomes a subhalo, sinking towards the inner region of
the host, its dark matter component could be significantly stripped
due to tidal destruction, leaving behind a baryon-dominated
overdensity. This is because the latter is much more concentrated than
the former and thus less prone to tidal stripping. Such baryonic
substructures ($10^{7\sim9}M_{\odot}$) may follow a similar or an even
more concentrated spatial distribution compared to their CDM subhalo
counterparts; their density profiles (being more compact) differ
completely from the latter due to the different nature of baryons and
dark matter. These surviving baryonic substructures of a similar mass
range as their CDM counterparts may also induce significant density
perturbations and thus cause radio flux anomalies. Interestingly
\citet{ShinEvans2008} investigated the possibility of using the
Milky-Way satellite galaxy population to explain the observed flux
anomaly frequencies. They found that the results strongly depend on
the applied density profiles (of the baryonic substructures); a
central density enhancement relative to the Milky-Way satellite
population of a factor of $10-100$ is needed in order to explain
observations.
% {\it We would like to leave the effects of baryonic substructures in
%  a forthcoming paper.}

Other possible sources of radio flux anomalies also include
oversimplified/improper lens modelling and radio propagation effects,
for which there is already evidence in a few systems (e.g., B0128+437,
B0712+472, B1555+375), and might have affected the model-predicted
flux ratios therein.
%It is therefore also crucial to establish whether the anomaly problem
%actually lies in the commonly applied method for lens modelling.

\subsection{Summary}
Discrepancies between the observed and model-predicted flux ratios
that assume a smooth lens are seen in a number of radio lenses. The
most favoured interpretation of these anomalies is that CDM
substructures perturb the lens potentials and alter image
magnifications (and thus flux ratios). In this work we particularly
study the cusp and fold relations in quadruple lenses to see how the
flux ratios $\Rcusp$ and $\Rfold$ would be affected by CDM
substructures.

In the first part of this paper, we assume that general smooth lens
potentials can be modelled as isothermal ellipsoids with a wide range
of axis ratios, higher-order multipole perturbations and randomly
oriented external shear (Sect. 3.1). We then take two sets of
state-of-the-art high resolution CDM cosmological simulations: the
Aquarius suite of galactic halos and the Phoenix suite of cluster
halos whose subhalo populations were rescaled to those expected in
group-sized halos (Sect. 3.2). By ray-tracing through the combined
(and perturbed) lens potentials, we produce a large sample of
quadruply-imaged quasars lensed by massive elliptical galaxies, and
predict their flux ratio probability distributions.

We find that host mass rescaling indeed makes a difference in the
final $\Rcusp$ and $\Rfold$ probability distributions (see
Fig. \ref{fig:CuspViolationTotal}). The projected radial distribution
of the surface number density of subhalos, as well as their dependence
on host halo masses and redshifts, are given in
Fig. \ref{fig:SubProjDistribution} and
\ref{fig:SubSpatialMassRed}. The subhalo mass fraction at around one
Einstein radius increases by a factor of 3 from Milky Way-sized host
halos to group-sized host halos (Sect. 3.2.2). As a result, using
subhalo populations in group-sized halos markedly increases the flux
anomaly frequencies compared to using those from Milky Way-sized halos
(Sect. 3.4). The forecasts as shown in Fig. 4 also clearly confirm
that systems which are more likely to show signatures of CDM
substructures through the induced anomalous flux ratios are those with
small image opening angle and/or image separation, or in other words,
highly magnified systems.

%The generic flux ratio probability distributions
%(Fig. \ref{fig:CuspViolationTotal}) clearly reflect the effects of
%density perturbations on flux ratios. As the magnification and local
%convergence follow $\mu\approx(1-2\kappa)^{-1}$, and thus
%$\delta\mu/\mu\propto\mu \delta\kappa$. A tiny density fluctuation
%near an image position that is close to the critical curve (where
%$\mu\rightarrow\infty$) will significantly perturb the local image
%magnification; while a density fluctuation around an image that is
%further away from the critical curve will be far less efficient in
%altering the image magnification via density perturbation. For this
%reason, image triplets and pairs with small separation are best probes
%of density perturbations (substructures) in lensing galaxies.

In the second part of this paper, we present results of case studies
for observed lens systems, all of which have radio measurements for
both cusp and fold relations. In these calculations we take the
best-fitting macroscopic lens models (Sect. 4.1), populating the
rescaled Aquarius and Phoenix subhalo populations above
$10^7h^{-1}M_{\odot}$ (due to resolution limit, see Sect. 4.2), and
subhalos with masses three orders of magnitudes lower
(Sect. 4.3). Through numerical experiments we confirm that
perturbation effects increase with increasing subhalo mass (assuming
point sources); but decrease with increasing size of a finite source
(Fig. \ref{fig:RcuspRfoldTestLens7}). We then study the probability
distributions of $\Rcusp$ and $\Rfold$ for mock samples that closely
resemble the specific image geometries in the observed systems,
predicting how likely it is to reproduce the measurements for each
system in presence of CDM subhalos (Sect. 4.6).

% We would like to point out that for systems with image triples/pairs
% of larger separation and lower magnifications (e.g., B1608 and B1933),
% we exclude them from our analysis. On the one hand, the currently
% applied lens modelling is too simple to be appropriate; on the other
% hand, their flux ratios are less susceptible to density perturbations.
%%the discrepancies between measurements and model predictions are more
%%likely attributed to the propagation effect in the interstellar medium
%%or to inaccuracies in simplified smooth lens models.

Focusing on those systems with closely located image triplets/pairs,
as can be seen from Fig. \ref{fig:ViolationIndivSystems1}, we find
that to have $\Rcusp$ and $\Rfold$ larger than the observed values the
probabilities are only $1\%-4\%$ for most systems. Only for
MG0414+114, a probability of $5\%-20\%$ is obtained. We conclude that
CDM substructures may not be the entire reason for the radio flux
anomaly problem; other sources, e.g., propagation effects and/or
inadequate lens modelling, could also be at work.  Apart from those,
baryonic (sub)structures with masses ranging from $10^6M_{\odot}$ to
$10^9M_{\odot}$ that survive the tidal destruction during galaxy
merger and accretion could also be important sources of density
fluctuation and thus (radio) flux ratio anomalies.

Comparisons between the results from two different methodologies
performed in this paper as well as in existing literature on flux
ratio anomalies (see Sect. 1) suggest that a proper study of flux
ratio anomalies needs a well-controlled sample of
lenses. Alternatively, investigations based on individual systems may
critically depend on the choice of the reference macroscopic model to
which substructures are added. More specifically, the nearby lens
environment can modify $\Rcusp$ (and $\Rfold$) at a level of tenths of
percents; variation in ellipticity and deviations from perfect
ellipses in the lens mass distribution (parametrized with multipole
terms) can also lead to significant changes of the flux ratios.
%The latter component is however impossible to evaluate independently
%%of the flux ratios and hence should be treated statistically.
Therefore, apart from local density perturbations, simplified lens
modelling which does not take into account the ingredients above can
also lead to a spurious mismatch between the observed and the
predicted flux ratios.

To make further progress on this problem, on the one hand more
detailed observations, e.g., higher resolution and deeper image and
spectroscopic data of the quadruple systems are needed to allow better
characterization and quantification of macro lens models; on the other
hand, high resolution hydrodynamic simulations that follow the
evolution of baryons as well as the interplay between baryons and dark
matter are necessary to assist in identifying all true culprits for
the radio flux-ratio anomalies.

\section*{ACKNOWLEDGEMENTS}
The authors would like to thank Matthias Bartelmann, Olaf Wucknitz,
Ben Metcalf, Simona Vegetti, Neal Jackson, Leon Koopmans and Caina Hao
for insightful discussion, and thank Aaron Ludlow for providing test
data and useful comments. This work was carried out partly on the
DiRAC-1 and DiRAC-2 supercomputing facilities at Durham University,
and partly on the Magny supercomputing facilities at HITS. DDX
acknowledges the Alexander von Humboldt foundation for the
fellowship. DS is supported by the German \emph{Deut\-sche
  For\-schungs\-ge\-mein\-schaft, DFG\/} project No. SL172/1-1, and by
a {\it {Back to Belgium}} grant from the Belgian Federal Science
Policy (BELSPO). This project is supported by the Strategic Priority
Research Programme ``The Emergence of Cosmological Structures'' of the
Chinese Academy of Sciences Grant No. XDB09000000 (LG, JW, and SM). LG
acknowledges supports from the 100-talents program of the Chinese
academy of science (CAS), the National Basic Research Program of China
- Program 973 under grant No. 2009CB24901, the {\small NSFC} grants
No. 11133003, the {\small MPG} Partner Group Family and the {\small
  STFC} Advanced Fellowship, and thanks the hospitality of the
Institute for Computational Cosmology (ICC) at Durham University. JW
acknowledges supports from the Newton Alumni Fellowship, the
1000-young talents program, the 973 program grant No.  2013CB837900,
2015CB857005, the CAS grant No. KJZD-EW-T01, and the NSFC grant
No. 11373029, 11390372, 11261140641.
%JW acknowledges supports from the Newton Alumni Fellowship, the
%1000-young talents program, the {\small CMST} grant No. 2013CB837900,
%the {\small NSFC} grant No. 11261140641, and the CAS grant
%No. KJZD-EW-T01. 
CSF acknowledges an ERC Advanced Investigator grant (COSMIWAY). SM
thanks the CAS and National Astronomical Observatories of CAS (NAOC)
for financial support. VS acknowledges financial support from the
Deutsche Forschungsgemeinschaft through Transregio 33, 'The Dark
Universe'. Phoenix is a project of the Virgo Consortium. Most
simulations were carried out on the Lenova Deepcomp7000 supercomputer
of the super Computing Center of CAS in Beijing. This work was also
supported in part by an STFC Rolling Grant to the ICC.

\onecolumn 
\appendix

\section{Summary of the best flux ratios for the sample of lensed systems}

We provide in Table ~\ref{tab:ObsSample} the best available flux
ratio measurements for the sample of lenses studied in the main
text. When flux ratios vary with spatial resolution due to resolved
structures in images, we provide measurements obtained at different
spatial resolution.
%We use however, when possible, flux ratios obtained with VLBI or
%VLBA observations (i.e. higher spatial resolution frames).
When available, we also report flux ratios averaged over several
epochs or corrected for time delays between images. In Table
~\ref{tab:ObsSample}, VLBA and VLBI images have typical beam sizes of
2 sq. ${\rm mas}$ while VLA and MERLIN frames have typical beam sizes
of 50 sq. ${\rm mas}$.

\begin{table*}
\centering \caption{Observed lenses with measurements of $\Rcusp$
and $\Rfold$ for the close triple images:} \label{tab:ObsSample}
\begin{minipage} {\textwidth}
\begin{tabular}[b]{l|ccccccccccccccc}\hline\hline

ID & Observation & $F_1$ & $F_2$ & $F_3$ & $\Rcusp$ & ${\Rfold}$ & Images & References \\
B0128$^\dagger$ & VLA 5 GHz 41 epochs    &  0.584$\pm$0.029 & 1.0$\pm$0.0      &  0.506$\pm$0.032 & $-$0.043$\pm$0.020 & 0.263$\pm$0.014 & B*-A-D*  & 1 \\
        & VLBA 5 GHz               &  2.8$\pm$0.28 & 10.6$\pm$1.06       &  4.8$\pm$0.48    & $-$0.165$\pm$0.055 & 0.582$\pm$0.034 & -   & 2 \\
        & Merlin 5 GHz             &  9.5$\pm$1   & 18.9$\pm$1           &  9.2$\pm$1       & $-$0.005$\pm$0.046 & 0.331$\pm$0.033 & -      & 3 \\
MG0414  & VLBI 8.5 GHz core        & 115.6$\pm$11.56  &  97$\pm$9.7      & 34$\pm$3.4           & 0.213$\pm$0.049 & 0.087$\pm$0.065 & A1-A2*-B  & 4 \\
        & VLA 15 GHz 4 epochs    & 157.0$\pm$5.5    &  138.75$\pm$5    & 138.75$\pm$2.25      & 0.361$\pm$0.012 & 0.062$\pm$0.024 & -      & 5 \\
        & MIR                     & 1.0$\pm$0.0      &  0.9$\pm$0.04    & 0.36$\pm$0.02        & 0.204$\pm$0.016 & 0.053$\pm$0.020 & -      & 6 \\
B0712   & VLA 5 GHz 41 epoch    & 1.0$\pm$0.0      &  0.843$\pm$0.061 & 0.418$\pm$0.037      & 0.254$\pm$0.024 & 0.085$\pm$0.030 & A-B*-C  & 1 \\
        & VLBA 5 GHz              & 10.7$\pm$0.15    &  8.8$\pm$0.15    & 3.6$\pm$0.15         & 0.238$\pm$0.009 & 0.097$\pm$0.010 & -    & 7 \\
B1422   & VLA 5 GHz 41 epochs      & 1.0$\pm$0.0      &  1.062$\pm$0.009 & 0.551$\pm$0.007      & 0.187$\pm$0.004 & $-$0.030$\pm$0.004 & A-B*-C & 1 \\
        & VLBA 8.4 GHz            & 152$\pm$2        &  164$\pm$2       & 81$\pm$1             & 0.174$\pm$0.006 & $-$0.038$\pm$0.009 & -   & 8 \\
B1555   & VLA 5 GHz 41 epochs      & 1.0$\pm$0.0      &  0.62$\pm$0.059  & 0.507$\pm$0.073      & 0.417$\pm$0.026 & 0.235$\pm$0.028  & A-B*-C   & 1 \\
B1608$^{\dagger\dagger}$  & VLA 8.5 GHz   & 2.045$\pm$0.01   &  1.037$\pm$0.01  & 1.0$\pm$0.001        & 0.492$\pm$0.002 & 0.327$\pm$0.003  & A-C*-B   & 9 \\
B1933$^\dagger$   & VLBA 5 GHz    &  4.7$\pm$0.4 & 19.4$\pm$0.4         &  5.4$\pm$0.4     & 0.315$\pm$0.016 & 0.610$\pm$0.009 & 3*-4-6* & 10 \\
        & VLA 15 GHz               &  2.5$\pm$0.4 & 15.5$\pm$0.4         &  3.2$\pm$0.4     & 0.462$\pm$0.018 & 0.722$\pm$0.009 & -      & 10 \\
B2045   & VLA 5 GHz 41 epochs    & 1.0$\pm$0.0      &  0.578$\pm$0.059 & 0.739$\pm$0.073      & 0.501$\pm$0.020 & 0.267$\pm$0.027  & A-B*-C & 1 \\
        & VLBA 5 GHz               & 1.0$\pm$0.01     &  0.61$\pm$0.01   & 0.93$\pm$0.01        & 0.520$\pm$0.003 & 0.242$\pm$0.007  & - & 11 \\
\hline
\end{tabular}
\\ Notes: the fluxes and errors (in Col. 3, 4 and 5) are directly
taken from the literature in their original units. When flux errors
are not available, we take 10\% of the measured fluxes as their
uncertainties. Image names (in Col. 8) associated with * indicate the
images with negative parities. ($\dagger$) Flux ratios are likely
affected by systematic errors due to scattering. ($\dagger\dagger$)
Quoted fluxes are after correction for the time delays.  References
(1) Koopmans et al. 2003; (2) Biggs et al. 2004 (Table 3); (3)
Phillips et al. 2000; (4) Ros et al. 2000; (5) Lawrence et al. 1995;
(6) Minezaki et al. 2009; (7) Jackson et al. 2000; (8) Patnaik et al.
1999; (9) Fassnacht et al. 1999; (10) Sykes et al. 1998; (11) McKean
et al. 2007.
\end{minipage}

\end{table*}

\section{Generalised isothermal lens with multipole perturbation and external shear}

Consider a lens potential composed of a singular isothermal
ellipsoidal, $m^{\rm th}$-mode multipole perturbation and external
shear:
\begin{equation}
\psi(\theta,\phi)=\psi_{\rm
SIE}(\theta,\phi)+\psi_m(\theta,\phi)+\psi_{\rm ext}(\theta,\phi),
\label{eq:lenspototsum}
\end{equation}
where $\theta$ and $\phi$ are the image position
$\vec{\theta}$=($\theta_{x}$, $\theta_{y}$) in polar coordinate:
$\theta=\sqrt{\theta_{x}^2+\theta_{y}^2}$ and
$\phi=\tan^{-1}(\theta_{y}/\theta_{x})$; $\psi_{\rm SIE}$, $\psi_m$
and $\psi_{\rm ext}$ are lens potentials of an singular isothermal
ellipsoidal, $m^{\rm th}$-mode multipole perturbation and external
shear, respectively. 
%The deflection angles and second-order derivatives of the total lens
%potential are then given by:
%\begin{equation}
%\left\{\begin{array}{l}
%\alpha_x(\theta,\phi)\equiv\frac{\partial\psi}{\partial \theta_x}=
%\alpha_{\textrm{SIE},x}(\theta,\phi)+\alpha_{m,x}(\theta,\phi)
%+\alpha_{\textrm{ext},x}(\theta,\phi) \\~~\\ % \nonumber \\
%\alpha_y(\theta,\phi)\equiv\frac{\partial\psi}{\partial \theta_y}=
%\alpha_{\textrm{SIE},y}(\theta,\phi)+\alpha_{m,y}(\theta,\phi)
%+\alpha_{\textrm{ext},y}(\theta,\phi) \\~~\\ %  \nonumber \\
%\psi_{11}(\theta,\phi)\equiv\frac{\partial^2\psi}{\partial
%\theta_x^2}=\psi_{\textrm{SIE},11}(\theta,\phi)+\psi_{m,11}
%(\theta,\phi)+\psi_{\textrm{ext},11} (\theta,\phi) \\~~\\ % \\
%\psi_{22}(\theta,\phi)\equiv\frac{\partial^2\psi}{\partial
%\theta_y^2}=\psi_{\textrm{SIE},22}(\theta,\phi)+\psi_{m,22}
%(\theta,\phi)+\psi_{\textrm{ext},22}(\theta,\phi) \\~~\\ %  \nonumber \\
%\psi_{12}(\theta,\phi)\equiv\frac{\partial^2\psi}{\partial
%\theta_x\theta_y}=\psi_{\textrm{SIE},12}(\theta,\phi)+\psi_{m,12}
%(\theta,\phi)+\psi_{\textrm{ext},12}(\theta,\phi). % \nonumber
%\label{eq:firstsecondpotsum}
%\end{array}\right.
%\end{equation}
In our numerical approach for lensing calculations, we tabulate to a
Cartesian mesh ($\theta_x$, $\theta_y$) in the image plane values of
the reduced deflection angle and second-order derivatives of the lens
potential. %Below we give the exact analytical formulae for the lensing
%quantities in Eq. (\ref{eq:firstsecondpotsum}). \\

%\subsection{Singular isothermal ellipsoidal plus $m^{\rm th}$-mode multipole perturbation}
For a generalised isothermal lens (plus perturbations), the lens
potential $\psi$ and convergence $\kappa$ follow the pair of equations
below (Keeton et al. 2003, Appendix B2):
\begin{equation}
\begin{array}{l} 
\displaystyle \psi(\theta,\phi)=\theta F(\phi) = \theta
\left[F_{\rm SIE}(\phi)+\sum\limits_{m=3,4}F_m(\phi)\right],\\ 
\displaystyle \kappa(\theta,\phi)=R(\phi)(2\theta)^{-1}=
\left[R_{\rm SIE}(\phi)+\sum\limits_{m=3,4}\delta
R_m(\phi)\right](2\theta)^{-1}. \label{eq:GIEmPair}
\end{array}
\end{equation}
From the Poisson equation $\nabla^2\psi=2\kappa$, $F(\phi)$ and
$R(\phi)$ are related by: $R(\phi)=F(\phi)+F^{\prime\prime}(\phi)$.
$F_{\rm SIE}(\phi)$ and $R_{\rm SIE}(\phi)$ are shape functions of a
singular isothermal ellipsoidal lens, while $F_m(\phi)$ and $\delta
R_m(\phi)$ describe the higher-order multipole perturbations. For
the generic lens model used in this work, only $m=3$ and 4 are
considered.

\subsection{Singular isothermal ellipsoid}
Specifically, if the isothermal ellipsoid's major and minor axes
coincide with the Cartesian axes, then the shape functions are given
by (\citealt{KassiolaKovner1993, KSB1994, KeetonKochanek1998}):
\begin{equation}
\begin{array}{l} 
\displaystyle R_{\rm SIE}(\phi) =
\frac{\Rein}{\sqrt{1-\epsilon\cos2\phi}},\\ 
\displaystyle F_{\rm SIE}(\phi)=\frac{\Rein}{\sqrt{2\epsilon}}\left[\cos\phi\tan^{-1}
\left(\frac{\sqrt{2\epsilon}\cos\phi}{\sqrt{1-\epsilon\cos2\phi}}\right)
+\sin\phi\tanh^{-1}\left(\frac{\sqrt{2\epsilon}\sin\phi}
{\sqrt{1-\epsilon\cos2\phi}}\right)\right]. \label{eq:SIERFpair}
\end{array}
\end{equation}
where $\Rein$ is the Einstein radius of the singular isothermal
ellipsoid, $\epsilon=(1-q^2)/(1+q^2)$ and $q\in(0,1]$ is the axis
  ratio of the ellipsoid. It can be shown that $R_{\rm SIE}(\phi)$ is
  the equation in polar coordinates of the ellipse at the critical
  curve, where $\kappa_{\rm SIE}=\frac{1}{2}$. $R_{\rm SIE}(\phi)$
  corresponds to the ellipse's equation in Cartesian coordinates:
\begin{equation}
\frac{\theta_{x}^2}{a^2}+\frac{\theta_{y}^2}{b^2} = 1,~
\textrm{where~~} a=\frac{\Rein}{\sqrt{1-\epsilon}}, ~
b=aq=\frac{\Rein}{\sqrt{1+\epsilon}}. \label{eq:isoSIECartesian}
\end{equation}
As convergence $\kappa_{\rm SIE}(\theta,\phi)=\frac{R_{\rm
SIE}(\phi)}{2\theta}$, the iso-$\kappa_{\rm SIE}$ contours follow
the ellipse $R_{\rm SIE}(\phi)$ and are scaled by $\theta^{-1}$. 

\subsection{Higher-order multipole perturbations}
Now consider adding a higher-order multipole perturbation $\delta
R_m(\phi)$ to the iso-$\kappa$ ellipse $R_{\rm SIE}(\phi)$, where
$\delta R_m(\phi)$ is defined as (see Keeton et al. 2003, Appendix
B2):
\begin{equation}
\delta R_m(\phi)=a_m\cos(m(\phi-\phi_m))
\label{eq:deltaRm}
\end{equation}
where $a_m$ ($>$0) and $\phi_m$ are the amplitude and
``orientation'' of the $m^{\rm th}$-order perturbation to the perfect
ellipse $R_{\rm SIE}(\phi)$.
%$\phi_m$ is measured counter-clockwise from the semi-major axis of the
%isothermal ellipsoidal.

In the particular case of the 4$^{\rm th}$-mode perturbation, an
elliptical galaxy would be more discy if $\phi_4=0$, and more boxy if
$\phi_4=\pi/4$ (which is the same as in the conventional definition
that $\delta R_4(\phi)=a_4\cos(4\phi)$, where $a_4>0$ corresponds to a
discy galaxy and $a_4<0$ corresponds to a boxy galaxy).

From Eq. (\ref{eq:GIEmPair}) it can be seen that, as the convergence
is $\kappa(\theta,\phi)=(R_{\rm SIE}(\phi)+\sum\limits_{m}\delta
R_m(\phi))(2\theta)^{-1}$, now the new iso-$\kappa$ contours follow
the perturbed ellipse $(R_{\rm SIE}(\phi)+\sum\limits_{m}\delta R_m(\phi))$ (at
$\kappa=\frac{1}{2}$) and are scaled by $\theta^{-1}$.

The corresponding shape function $F_m(\phi)$ is given by (see Keeton
et al. 2003):
\begin{equation}
F_m(\phi)=\frac{1}{1-m^2}a_m\cos(m(\phi-\phi_m)).
\label{eq:Fm}
\end{equation}

%The deflection angles and second order derivatives of the potential
%due to the $m^{\rm th}$ perturbation are given by:
%\begin{equation}
%\left\{\begin{array}{l}
%\alpha_{m,x}(\theta,\phi)=F_m(\phi)\cos\phi-F^{\prime}_m(\phi)\sin\phi
%\\ ~~ \\
%\alpha_{m,y}(\theta,\phi)=F_m(\phi)\sin\phi+F^{\prime}_m(\phi)\cos\phi
%\\ ~~ \\\psi_{m,11}(\theta,\phi)=[\theta]^{-1}\sin^2\phi[F_m(\phi)+
%F^{\prime\prime}_m(\phi)]=[\theta]^{-1}\sin^2\phi\delta R_m(\phi)
%\\ ~~ \\\psi_{m,22}(\theta,\phi)=[\theta]^{-1}\cos^2\phi[F_m(\phi)+
%F^{\prime\prime}_m(\phi)]=[\theta]^{-1}\cos^2\phi\delta
%R_m(\phi)\\~\\
%\psi_{m,12}(\theta,\phi)=-[2\theta]^{-1}\sin2\phi[F_m(\phi)+
%F^{\prime\prime}_m(\phi)]=-[2\theta]^{-1}\sin2\phi\delta R_m(\phi)
%\end{array}\right.
%\label{eq:alphaphi12Fm}
%\end{equation}
%where $F^{\prime}_m=\frac{\partial F_m}{\partial \phi}$ and
%$F^{\prime\prime}_m=\frac{\partial^2 F_m}{\partial \phi^2}$.\\

The physical quantity of $\delta R_m$ in Eq. (\ref{eq:deltaRm}) is the
same as in Hao et al. (2006), where the expression is given by:
\begin{equation}
\delta R_m(\phi)=\alpha_m\cos(m\phi)+\beta_m\sin(m\phi).
\label{eq:deltaRmHao}
\end{equation}
In their work, $\alpha_m/a$ and $\beta_m/a$ (where $a$ is the
semi-major axis length of the perfect ellipse) for $m=3,~4$, and
ellipticity $e(\equiv1-q)$ of the elliptical isophotes were measured
within the Petrosian half-light radii. We use these values in our
main lens modelling. Notice that Eq. (\ref{eq:deltaRmHao}) can also be
re-written as: $\delta
R_m(\phi)=\sqrt{\alpha_m^2+\beta_m^2}\cos(m(\phi-\phi_m))$, where
$\phi_m=\frac{1}{m}\tan^{-1}(\beta_m/\alpha_m)\in\frac{1}{m}[0,2\pi)$. Comparing
  with Eq. (\ref{eq:deltaRm}), it can be seen that:
\begin{equation}
\begin{array}{l}
\displaystyle
a_m=\sqrt{\alpha_m^2+\beta_m^2}\equiv\sqrt{(\alpha_m/a)^2+(\beta_m/a)^2}\times a_{\rm SIE},\\
\displaystyle \phi_m=\frac{1}{m}\tan^{-1}(\beta_m/\alpha_m).
\end{array}
\end{equation}
where $a_m$ is re-normalized at $\kappa=\frac{1}{2}$;
$a_{\rm SIE}=\frac{\Rein}{\sqrt{1-\epsilon}}$ as given in
Eq. (\ref{eq:isoSIECartesian}). 
%Therefore, the model parameters for higher-order perturbations
%($m=3,~4$) can be fixed using observational samples (Hao et al. 2006).

\subsection{Constant external shear}
The lens potential $\psi^{\rm ext}(\theta,\phi)$ caused by a
constant external shear is given by:
\begin{equation}
\psi_{\rm ext}(\theta,\phi)=-\frac{\gamma_{\rm
ext}}{2}\theta^2\cos(2(\phi-\phi_{\rm ext})),
\end{equation}
where $\gamma_{\rm ext}(>0)$ is the shear amplitude and $\phi_{\rm
  ext}\in[0,\pi)$ is the position angle of the shear mass, measured
  counter-clockwise from the semi-major axis of the isothermal
  ellipsoid. External shear will not contribute to external
  convergence, i.e., $\kappa_{\rm ext}=0$.
%The deflection angle and second-order derivatives are given by:
%\begin{equation}
%\left\{\begin{array}{l}
%\alpha_{\textrm{ext},x}(\theta,\phi)=-\theta\gamma_{\rm
%ext}\cos(\phi-2\phi_{\rm ext})
%\\~~\\ \alpha_{\textrm{ext},y}(\theta,\phi)=\theta\gamma_{\rm
%ext}\sin(\phi-2\phi_{\rm ext}) \\~~\\
%\psi_{\textrm{ext},11}(\theta,\phi)=-\gamma_{\rm ext}\cos(2\phi_{\rm ext}) \\~~\\
%\psi_{\textrm{ext},22}(\theta,\phi)=\gamma_{\rm ext}\cos(2\phi_{\rm ext}) \\~~\\
%\psi_{\textrm{ext},12}(\theta,\phi)=-\gamma_{\rm ext}\sin(2\phi_{\rm ext}).
%\end{array}\right.
%\end{equation}
In this work, we assume random external shear orientation in each
simulated lens system.

\twocolumn
\bibliographystyle{mn2e}
%\bibliography{ms_xudd}
\label{lastpage}

\end{document}